\documentclass[%
 reprint,
superscriptaddress,
longbibliography,
 amsmath,amssymb,
 prl,
 citeautoscript,
]{revtex4-1}

\usepackage{graphicx}
\usepackage{dcolumn}
\usepackage{bm}
\usepackage{amsmath}
\usepackage{amssymb}
\usepackage{amsthm}
\usepackage{units}
\usepackage{gensymb}
\usepackage{xcolor}

\begin{document}

\title{Odd mobility of a passive tracer in a chiral active fluid}

\author{Anthony R. Poggioli}
\email{arpoggioli@berkeley.edu}
\affiliation{Department of Chemistry, University of California, Berkeley}
\affiliation{Kavli Energy NanoScience Institute, Berkeley, California}

\author{David T. Limmer}
\email{dlimmer@berkeley.edu}
\affiliation{Department of Chemistry, University of California, Berkeley}
\affiliation{Kavli Energy NanoScience Institute, Berkeley, California}
\affiliation{Materials Science Division, Lawrence Berkeley National Laboratory}
\affiliation{Chemical Science Division, Lawrence Berkeley National Laboratory}

\date{\today}

\begin{abstract}
Chiral active fluids break both time-reversal and parity symmetry, leading to exotic transport phenomena unobservable in ordinary passive fluids. We develop a generalized Green-Kubo relation for the anomalous lift experienced by a passive tracer suspended in a two-dimensional chiral active fluid subjected to an applied force. This anomalous lift is characterized by a transport coefficient termed the odd mobility. We validate our generalized response theory using molecular dynamics simulations, and we show that the asymmetric tracer mobility may be understood mechanically in terms of asymmetric deformations of the tracer-fluid density distribution function. We show that the even and odd components of the mobility decay at different rates with tracer size, suggesting the possibility of size-based particle separation using a chiral active working fluid.
\end{abstract}
\maketitle

Chiral active fluids are composed of constituents that convert energy into directed rotational motion \cite{Banerjee, Cory_MD, Vitellis_pretty_pictures}. The quiescent state of such a fluid breaks both parity and time-reversal symmetry and accordingly is not a state of thermodynamic equilibrium. These broken symmetries lead to a variety of exotic transport phenomena \cite{Avron, Banerjee, waves_Kranthi, bubble_dynamics_odd_viscosity, Vitellis_pretty_pictures, Cory_MD, Cory_odd_diffusivity, Hosaka_mobility, orthogonal_fluid_motion, 3d_odd_viscosity, boundary_waves, crystal, cargo, Hall-like, free_surface}. For example, an odd viscosity that links dilational flow to shear stress and sheared flow to normal stress can translate forcing into fluid motion in an orthogonal direction \cite{Avron, orthogonal_fluid_motion,3d_odd_viscosity}. Describing transport phenomena in such nonequilibrium fluids is challenging because classical linear response theories are valid perturbatively only about an equilibrium steady state \cite{deGroot+Mazur}. Here, we consider a similar transport process, the motion of a passive tracer suspended in a two-dimensional, odd viscous chiral active fluid and subjected to a constant force, and we develop a response theory based on ensembles of trajectories \cite{Chloe1, Chloe2}. Recent work has examined the advective \citep{cargo} and driven \citep{Hosaka_mobility} transport of passive tracers in odd viscous fluids. These theoretical treatments are phenomenological, however, illustrating the need for a microscopic description of tracer mobility in chiral active fluids. The framework employed here can be extended to arbitrary order in the applied forcing and is valid for a reference state arbitrarily far from equilibrium \cite{review}.

In general, the velocity response of a passive tracer is characterized by a second rank tensor called the mobility tensor by the relation $\left\langle {\boldsymbol V} \right\rangle_{\boldsymbol F} = \boldsymbol{\mu} \cdot {\boldsymbol F}$, where ${\boldsymbol V}$ is the tracer velocity, $\boldsymbol{\mu}$ is the mobility tensor, ${\boldsymbol F}$ is an external force applied  to the tracer, and angled brackets denote a nonequilibrium ensemble average \cite{Cory_odd_diffusivity, Hosaka_mobility}. In passive fluids, isotropy requires that the mobility tensor be proportional to the unit tensor, introducing a single transport coefficient, $\mu_\mathrm{e}$, termed the even mobility. However, in two dimensional fluids breaking parity and time-reversal symmetry, the isotropy of the second rank two-dimensional Levi-Civita tensor permits the emergence of an additional transport coefficient, $\mu_\mathrm{o}$, termed the odd mobility, analogous to the emergence of the odd viscosity \cite{Avron}. The general expression for the mobility tensor is given in Cartesian coordinates by
\begin{equation}
\boldsymbol{\mu} = \mu_\mathrm{e} {\boldsymbol 1} + \mu_\mathrm{o} \boldsymbol{\epsilon} =
\begin{pmatrix}
\mu_\mathrm{e} & \mu_\mathrm{o} \\
-\mu_\mathrm{o} & \mu_\mathrm{e},
\end{pmatrix}
\label{eqn:mobility_tensor}
\end{equation}
where ${\boldsymbol 1}$ is the unit tensor, and $\boldsymbol{\epsilon}$ is the Levi-Civita tensor. Under a force ${\boldsymbol F} = F \hat{\boldsymbol x}$ applied in the positive $x$-direction, the resulting tracer velocity is then $\left\langle {\boldsymbol V} \right\rangle_{\boldsymbol F} = \mu_\mathrm{e} F \hat{\boldsymbol x} - \mu_\mathrm{o} F \hat{\boldsymbol y}$, indicating that the particle experiences an anomalous lift proportional to the odd mobility in the direction orthogonal to the applied force. This lift corresponds to the conversion of directed rotational motion of the underlying fluid into a uniform deflection of the tracer particle. Such a behavior could be observed experimentally by subjecting tracers in an odd viscous fluid \cite{free_surface} to an external force.

\begin{figure}[t]
	\centerline{\includegraphics[width=8.5cm]{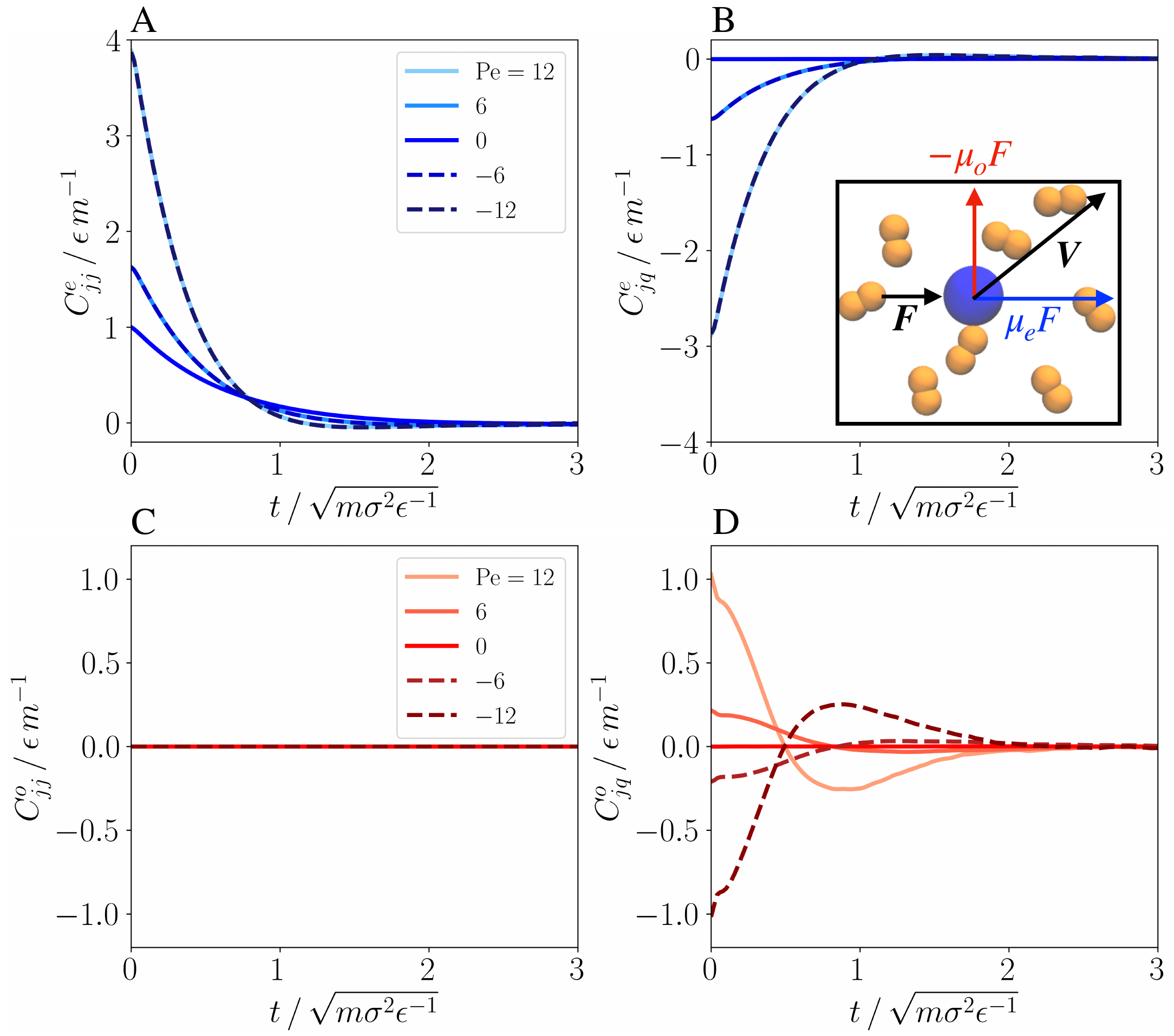}}
    \caption{Correlation functions contributing to the even and odd mobility response of a passive tracer for several different P\'{e}clet numbers and density $\bar{n} = 0.2 \, \sigma^{-2}$. A) The current autocorrelation for the even mobility. B) The current-frenesy cross-correlation for the even mobility. C) and D) The current autocorrelation (C) and current-frenesy cross-correlation for the odd mobility. The inset in panel B shows a simulation snapshot and the decomposition of the velocity response of a passive tracer subjected to an applied force in terms of the even and odd mobility coefficients. The blue particle is the passive tracer, and the orange particles are the actively rotating dimers. Errors are smaller than the line thickness.}
\label{fig:Fig1}
\end{figure}

Here, we consider odd mobility numerically with a simple model containing a passive Weeks-Chandler-Andersen (WCA) particle \cite{WCA} suspended in a fluid of active dimers \cite{Cory_MD}. The dimers are composed of monomers linked by stiff harmonic bonds that interact with each other and the passive tracer via a purely repulsive WCA potential characterized by an energy parameter $\beta \epsilon = 1$, where $\beta \equiv 1/k_B T$ is the inverse of the temperature times Boltzmann's constant. Each monomer is also subjected to an active force of constant magnitude $F_a$, directed perpendicular to the instantaneous dimer bond. This active force results in a directed active torque experienced by each dimer of average magnitude $F_a d_0$, where $d_0 = \sigma$ is the equilibrium bond length set equal to the monomer diameter $\sigma$. The active forces are oriented such that $F_a > 0$ corresponds to a counterclockwise active torque. The monomers undergo underdamped Langevin dynamics with the equation of motion
\begin{equation}
m \dot{\boldsymbol v}_i^{\alpha} = - \gamma {\boldsymbol v}_i^{\alpha} + {\boldsymbol F}_{{\rm c},i}^{\alpha} + {\boldsymbol F}_{a,i}^{\alpha} + \sqrt{2 \gamma /\beta} {\boldsymbol N}_i^{\alpha}(t),
\label{eqn:tracer_eom}
\end{equation}
where $i$ is the dimer index and $\alpha \in \left\{1,2\right\}$ indexes the monomers on dimer $i$, $m$ is monomer mass, ${\boldsymbol F}_{{\rm c},i}^{\alpha}$ are the conservative forces acting on monomer $i$-$\alpha$, $ {\boldsymbol F}_{a,i}^{\alpha}$ is the active force, and $\gamma$ is a linear friction coefficient describing dissipation from the particles to an underlying quiescent fluid bath. (See Supplemental Materials, including Refs. \cite{joanny, 2d_Stokes}.) ${\boldsymbol N}_i(t)$ is a delta-correlated Gaussian fluctuating force with mean zero and unit variance. A simulation snapshot is shown in the inset in Fig. \ref{fig:Fig1}A, and details of the dimer model can be found in Refs. \cite{Cory_MD, Cory_github}. This model represents a minimal model of a chiral active fluid exhibiting odd viscosity \cite{Cory_MD}. The Supplemental Materials contain further simulation details.

\begin{figure}[h!]
	\centerline{\includegraphics[width=8.5cm]{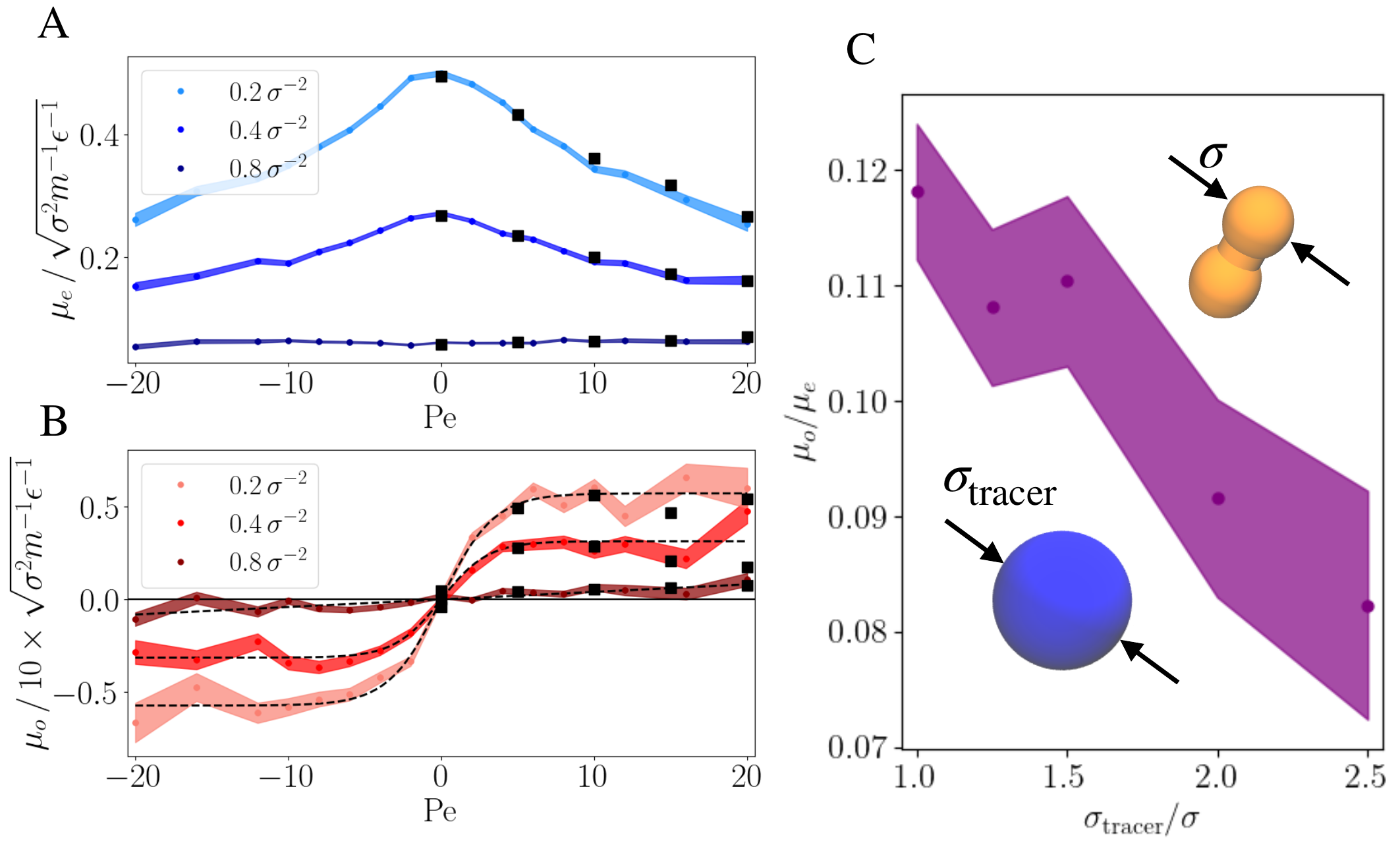}}
    \caption{Mobility as a function of activity, density, and tracer diameter. A) and B)  show even and odd mobility as a function of P\'{e}clet number for three different densities. C) Ratio of odd to even mobility as a function of tracer diameter for $\bar{n} = 0.2 \, \sigma^{-2}$. Shaded regions indicate standard errors. The black squares in panels A and B show the results of forced mobility calculations. The dashed black lines in panel B are guides to the eye.}
\label{fig:Fig2}
\end{figure}

The tracer equation of motion is
\begin{equation}
m \dot{\boldsymbol V} = - \gamma {\boldsymbol V} + {\boldsymbol F}_{\rm WCA} + {\boldsymbol F} + \sqrt{2 \gamma /\beta} {\boldsymbol N}(t),
\label{eqn:tracer_eom}
\end{equation}
where ${\boldsymbol V} \equiv \dot{\boldsymbol X}$ is the tracer velocity, ${\boldsymbol X}$ is the tracer position, ${\boldsymbol F}_{\rm WCA}$ is the total WCA force on the tracer from the surrounding monomers, and ${\boldsymbol F}$ is a constant external force. We take the tracer mass and friction coefficient equal to that for the monomers in all simulations. Except in Fig. \ref{fig:Fig2}C, we also take the tracer diameter $\sigma_{\rm tracer}$ to be equal to $\sigma$. The dimer activity has a profound influence on the tracer dynamics. This effect is contained in the average structure of ${\boldsymbol F}_{\rm WCA}$, as we will explore below.

The Gaussian nature of the noise allows us to write the path probability density as \cite{cugliandolo2019building}
\begin{equation}
\mathbb{P}_{\boldsymbol F} \left[ {\rm X} \right] \propto e^{- \beta S_{\boldsymbol F} \left[ {\rm X} \right]},
\label{eqn:path_probability}
\end{equation}
where ${\rm X} \equiv \left\{ {\boldsymbol X}(t) \right\}_{t = 0}^{t_N}$ is a tracer trajectory of duration $t_N \rightarrow \infty$, and the path action under an applied force ${\boldsymbol F}$ is given by the Onsager-Machlup construction \cite{Onsager+Machlup1, Onsager+Machlup2} as
\begin{equation}
\begin{split}
S_{\boldsymbol F} \left[ {\rm X} \right] &= \frac{1}{4\gamma} \int_0^{t_N} {\rm d}t \, \left( m \dot{\boldsymbol V} + \gamma {\boldsymbol V} - {\boldsymbol F}_{\rm WCA} - {\boldsymbol F} \right)^2 \\
&\equiv S_0 \left[ {\rm X} \right] + \Delta S_{\boldsymbol F} \left[ {\rm X} \right].
\end{split}
\label{eqn:path_action}
\end{equation}
In the second equality we have extracted the unbiased path action $S_0 \left[ {\rm X} \right]$ obtained when ${\boldsymbol F} = 0$. The relative path action $\Delta S_{\boldsymbol F} \left[ {\rm X} \right]$ is defined by
\begin{equation}
\frac{\mathbb{P}_{\boldsymbol F} \left[ {\rm X} \right]}{\mathbb{P}_0 \left[ {\rm X} \right]}= e^{+\beta \Delta S_{\boldsymbol F} \left[ {\rm X} \right]},
\label{eqn:path_reweighting}
\end{equation}
with
\begin{equation}
\begin{split}
\Delta S_{\boldsymbol F} \left[ {\rm X} \right] &\approx
\frac{1}{2} {\boldsymbol J} \cdot {\boldsymbol F} + \frac{1}{2} {\boldsymbol Q} \cdot {\boldsymbol F},
\end{split}
\label{eqn:relative_path_action}
\end{equation}
where we have defined the time-extensive current ${\boldsymbol J} \equiv \int {\rm d}t \, {\boldsymbol V}$ and frenesy ${\boldsymbol Q} \equiv - \int {\rm d}t \, \left( {\boldsymbol F}_{\rm WCA} / \gamma \right)$ \cite{basu2015nonequilibrium}. We have retained terms only to first order in ${\boldsymbol F}$ as we are interested only in the linear response, and we have dropped the temporal boundary term proportional to $\int {\rm d} t \, m \dot{\boldsymbol V}$ because its contribution to the mobility exactly vanishes in the limit $t_N \rightarrow \infty$. (See Supplemental Materials.)

Eq. \ref{eqn:path_reweighting} allows us to reweight path ensemble averages taken in a forced (${\boldsymbol F} \neq 0$) ensemble to those in an unforced one, which we may use to develop a linear response expression for the mobility. Rewriting the trajectory average of ${\boldsymbol J}$ in the presence of ${\boldsymbol F}$ as one in its absence using Eq.~\ref{eqn:path_reweighting} and expanding to first order in ${\boldsymbol F}$, we obtain
\begin{equation}
\left\langle \delta {\boldsymbol J} \right\rangle_{\boldsymbol F} = t_N \left\langle {\boldsymbol V} \right\rangle_{\boldsymbol F} = \frac{\beta}{2} \left( \left\langle \delta {\boldsymbol J} \otimes \delta {\boldsymbol J} \right\rangle_0 + \left\langle \delta {\boldsymbol J} \otimes \delta {\boldsymbol Q} \right\rangle_0 \right) \cdot {\boldsymbol F},
\label{eqn:linear_response1}
\end{equation}
where $\delta \mathcal{O} \equiv \mathcal{O} - \left\langle \mathcal{O} \right\rangle_0$. We insert our definition for the mobility tensor $\left\langle {\boldsymbol V} \right\rangle = {\boldsymbol \mu} \cdot {\boldsymbol F}$ and differentiate with respect to ${\boldsymbol F}$ to obtain a linear response relationship for the mobility tensor directly:
\begin{equation}
\boldsymbol{\mu} = \frac{\beta}{2 t_N} \left( \left\langle \delta {\boldsymbol J} \otimes \delta {\boldsymbol J} \right\rangle_0 + \left\langle \delta {\boldsymbol J} \otimes \delta {\boldsymbol Q} \right\rangle_0 \right).
\label{eqn:linear_response2}
\end{equation}
This equation represents the key theoretical result of this work. The first term on the right hand side of Eq. \ref{eqn:linear_response2} contains only the steady state fluctuations in the current and corresponds to the ordinary `thermodynamic' response obtained about equilibrium steady states. The second term contains the current-frenesy correlation and is nonvanishing only in nonequilibrium reference states. Physically, the frenesy represents microscopic motion unassociated with the current. Though the particular form of the frenesy depends on the nature of the underlying microscopic dynamics \cite{Maes_frenesy,Chloe3_PDF,Chloe4_PDF}, for tracer motion well described by the underdamped Langevin equation given in Eq. \ref{eqn:tracer_eom}, the frenesy requires knowledge only of the forces and their time correlations. It therefore in principle may be measured experimentally for such a system.

We can make explicit the time-integrals appearing in the definitions of ${\boldsymbol J}$ and ${\boldsymbol Q}$ in Eq. \ref{eqn:linear_response2}. Using time-translation invariance and averaging the diagonal and off-diagonal components of the mobility tensor, we find
\begin{equation}
\mu_\mathrm{e} = \beta \int_0^{\infty} {\rm d} t \, \left[ C_{jj}^\mathrm{e}(t) + C_{jq}^\mathrm{e}(t) \right],
\label{eqn:mue}
\end{equation}
and
\begin{equation}
\mu_\mathrm{o} = \beta \int_0^{\infty} {\rm d} t \, C_{jq}^\mathrm{o}(t),
\label{eqn:muo}
\end{equation}
with
\begin{equation}
C_{jj}^\mathrm{e}(t) \equiv \frac{1}{2}\sum_{\alpha=x,y}  \left\langle \delta j_\alpha(t) \delta j_\alpha (0) \right\rangle_0,
\label{eqn:Cjje}
\end{equation}
\begin{equation}
C_{jq}^\mathrm{e}(t) \equiv \frac{1}{4} \sum_{\alpha=x,y} \left[ \left\langle \delta j_\alpha(t) \delta q_\alpha (0) \right\rangle_0 + \left\langle \delta q_\alpha(t) \delta j_\alpha (0) \right\rangle_0 \right] \, ,
\label{eqn:Cjqe}
\end{equation}
and
\begin{eqnarray}
C_{jq}^\mathrm{o}(t) \equiv \frac{1}{4} && \left[ \left\langle \delta j_x(t) \delta q_y (0) \right\rangle_0 + \left\langle \delta q_y(t) \delta j_x (0) \right\rangle_0 \right] \\
&& - \frac{1}{4} \left[ \left\langle \delta j_y(t) \delta q_x (0) \right\rangle_0 + \left\langle \delta q_x(t) \delta j_y (0) \right\rangle_0 \right] \nonumber .
\label{eqn:Cjqo}
\end{eqnarray}
In the above, ${\boldsymbol j} \equiv {\boldsymbol V}$, and ${\boldsymbol q} \equiv -{\boldsymbol F}_{\rm WCA}/\gamma$. The contribution of the current autocorrelation to the odd mobility exactly vanishes. This can be seen clearly from Eq. \ref{eqn:linear_response2} where the thermodynamic contribution to the mobility is proportional to the explicitly symmetric variance of the time extensive current.

Fig. \ref{fig:Fig1} shows the contributions of each correlation function to the mobilities. We plot the results for several different activities, quantified by a P\'{e}clet number ${\rm Pe} \equiv 2 \beta F_a d_0$. We include the results for equilibrium, ${\rm Pe} = 0$, for which the odd mobility must vanish. Indeed, we observe that $C_{jq}^\mathrm{o}$ is zero. For finite ${\rm Pe}$, both the current (thermodynamic) and current-frenesy (frenetic) contributions to $\mu_e$ are invariant under change of sign of ${\rm Pe}$. This is a reflection of the fact that $\mu_\mathrm{e}$ enters into the dissipation, $\left\langle {\boldsymbol V} \right\rangle_{\boldsymbol F} \cdot {\boldsymbol F} = \mu_\mathrm{e} F^2$, and hence is constrained to remain positive under time-reversal. It therefore must be invariant under parity inversion, $F_a \rightarrow -F_a$, in order to preserve overall parity-time symmetry.

The initial values of the thermodynamic and frenetic correlation functions for the even mobility both grow with increasing magnitude ${\rm Pe}$, though the current variance grows more rapidly. Adding these curves shows that the growth of the frenetic contribution acts to maintain a constant initial value of the overall correlation function, while the relatively larger frenetic integral relaxation time results in a decreasing overall relaxation time with increasing activity, and hence a decreasing even mobility. (See Supplemental Materials.) This is consistent with the decrease in overall even mobility with increasing activity shown in Fig. \ref{fig:Fig2}A.

The thermodynamic contribution to the odd mobility, $C_{jj}^\mathrm{o}$, exactly vanishes for all activities. As noted above this is predicted analytically and reflects the fact that the odd mobility is a purely nonequilibrium phenomenon. The frenetic correlation functions show a rich structure, possessing an initial shift in their decay rate at short times and a change in sign at longer times. The curves are inverted under a change in sign of $F_a$, reflecting the parity -- and hence time-inversion -- antisymmetry of the odd mobility. This is permitted because the odd mobility does not contribute directly to the tracer dissipation, analogous to the odd viscosity of the underlying fluid \cite{Avron}.

In Fig. \ref{fig:Fig2}, we show the even and odd mobility responses as measured by Eq. \ref{eqn:linear_response2} as a function of activity for several different densities. As anticipated, both the even and odd mobility responses are suppressed as density is increased. The maximum ratio of odd to even mobilities is obtained for the lowest density examined, $\bar{n} = 0.2 \, \sigma^{-2}$, and is roughly $20 \%$ for $\left| {\rm Pe} \right| = 20$. Surprisingly, we find that the even mobility is maximized for ${\rm Pe} = 0$ and decreases with increasing activity, demonstrating that a simple understanding of the activity in terms of an effective temperature does not hold in this system, as previous results have reported an increase in the mean squared diffusivity with ${\rm Pe}$ \cite{Cory_odd_diffusivity}.

At the highest density, $\bar{n} = 0.8 \, \sigma^{-2}$, the odd mobility is observed to vary linearly over the range of P\'{e}clet numbers examined, with positive values, corresponding to downward deflections under a force applied in the postive $x$-direction, obtained for counterclockwise dimer rotations and {\it vice verse}. However, we find that at lower densities the odd mobility appears to saturate to a constant value as $\left| {\rm Pe} \right|$ is increased. Combined with the observed monotonically decreasing nature of $\mu_\mathrm{e}$ with increasing $\left| {\rm Pe} \right|$, this indicates that the ratio of odd to even mobility may be made arbitrarily large for high activity, suggesting the possibility of a regime in which the particle is deflected by essentially $90^\circ$ from the applied force. We note that in Figs. \ref{fig:Fig2}A and B we have compared our simulation results in the reference steady state to runs in which a force is applied directly to the tracer particle and its velocity is directly measured, and the results are in excellent agreement.

Finally, in Fig. \ref{fig:Fig2}C, we plot the ratio of even to odd mobility as a function of tracer diameter, $\sigma_{\rm tracer}$. We plot only for a single ${\rm Pe}$ as we observe that this curve is essentially independent of ${\rm Pe}$ for finite activity. We find that the ratio of mobilities decreases with increasing $\sigma_{\rm tracer}$, indicating that $\mu_\mathrm{o}$ decreases more rapidly than $\mu_\mathrm{e}$ as tracer size is increased. The angle $\theta$ between applied force and resultant tracer velocity is given by ${\rm tan} \, \theta = - \mu_\mathrm{o} / \mu_\mathrm{e}$ \cite{Reichhardt1, Reichhardt2}, indicating that the particle deflection angle is therefore size-dependent. This suggests the possibility of size-based particle separation using an odd viscous fluid as a working fluid.

We can understand the origin of the trends in mobility mechanically by deriving a relationship between the tracer-monomer pair distribution function (PDF) and the mobility response \cite{Chloe3_PDF, Chloe4_PDF, Liao_PDFmechanics}. Averaging the tracer equation of motion, Eq. \ref{eqn:tracer_eom}, over the noise, we obtain
\begin{equation}
\left\langle {\boldsymbol V} \right\rangle_{\boldsymbol F} = \frac{1}{\gamma} \left( {\boldsymbol F} + \left\langle {\boldsymbol F}_{\rm WCA} \right\rangle_{\boldsymbol F} \right).
\label{eqn:average_tracer_eom}
\end{equation}
Given the pairwise nature of the repulsive interaction, the average of the components of the velocity parallel and orthogonal to the applied force can be written as
\begin{equation}
V_{||} = \frac{F}{\gamma} + \frac{\bar{n}}{\gamma} \int {\rm d} {\boldsymbol x} \, g\left( {\boldsymbol x} \left| {\boldsymbol F} \right. \right) F_{{\rm WCA},||} \left( {\boldsymbol x} \right),
\label{eqn:tangential_velocity}
\end{equation}
and
\begin{equation}
V_{\bot} = \frac{\bar{n}}{\gamma} \int {\rm d} {\boldsymbol x} \, g\left( {\boldsymbol x} \left| {\boldsymbol F} \right. \right) F_{{\rm WCA},\bot} \left( {\boldsymbol x} \right),
\label{eqn:orthogonal_velocity}
\end{equation}
respectively, where $g\left( {\boldsymbol x} \left| {\boldsymbol F} \right. \right)$ is the tracer-monomer PDF in the presence of an applied force ${\boldsymbol F}$ \cite{hansen2013theory, onsager1927report}, and ${\boldsymbol F}_{\rm WCA} \left( {\boldsymbol x} \right)$ is the force field characterizing the WCA interaction between monomers and the tracer particle held fixed at ${\boldsymbol X} = 0$.

\begin{figure}[t]
	\centerline{\includegraphics[width=8.5cm]{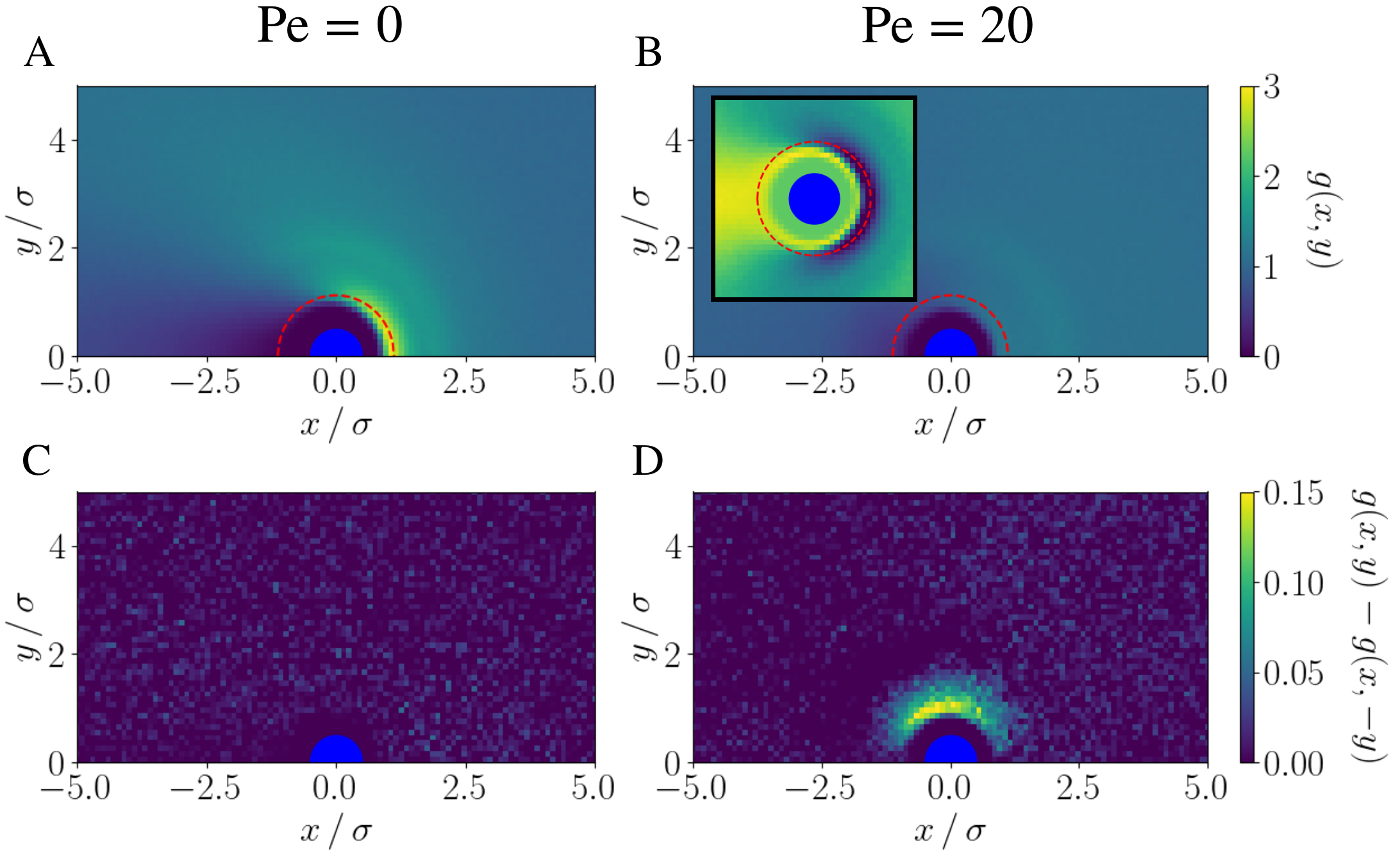}}
    \caption{Tracer-monomer pair distribution function (PDF) for a density of $\bar{n} = 0.2 \, \sigma^{-2}$ and different P\'{e}clet numbers under a force applied to the tracer in the positive $x$-direction of magnitude $F = 10 \, \epsilon \, \sigma^{-1}$. The left and right columns correspond to P\'{e}clet numbers of $0$ and $20$. The upper row shows the PDF in the upper half plane, while the lower row shows the PDF in the upper half plane minus the PDF in the lower half plane, illustrating the reflection asymmetry in $y$. The solid blue half circle indicates the location of the tracer particle. The inset in panel B shows the difference in PDFs obtained at ${\rm Pe} = 20$ and $0$, and the dashed red line in panels A and B and the inset indicates the cutoff of the WCA interaction.}
\label{fig:Fig3}
\end{figure}

We take ${\boldsymbol F}$ to be directed along the positive $x$-direction. Intuitively, we anticipate that a relative decrease in $V_{||}$ -- and hence $\mu_\mathrm{e}$ -- with increasing $\left| {\rm Pe} \right|$ will be associated with an increase in monomer density accumulated on the front of the particle. However, we observe the opposite in Figs. \ref{fig:Fig3}A and B, where we plot the PDF for ${\rm Pe} = 0$ and $20$ under an applied force. Instead, we observe that the ridge of monomer density accumulated on the front of the tracer decreases substantially in magnitude with increasing activity, ostensibly suggesting that $\mu_\mathrm{e}$ should increase with increasing magnitude ${\rm Pe}$ -- the opposite of what is observed in Fig. \ref{fig:Fig2}. This apparent inconsistency is resolved by considering the relative density distribution.

Though the accumulation of monomer density in  front of the tracer decreases with increasing activity, a high density ridge is pushed further into the interaction region with the tracer, resulting in a much stronger resistive interaction. This is clearly illustrated in the inset in Fig. \ref{fig:Fig3}B, which shows the PDF obtained for ${\rm Pe} = 20$ less that obtained for ${\rm Pe} = 0$. Direct integration confirms that this shift in particle distribution results in the suppression of the even mobility at finite activity. Interestingly, this shift in the high density ridge is not observed in the PDF calculated using the dimer center of mass coordinates, indicating that the dimers accumulated at the front of the tracer are preferentially oriented with a component of their bond vector orthogonal to the face of the tracer, and that this `orientational locking' increases with activity. This mechanism is fundamentally linked to the biased nature of the rotational motion of the dimers and is enhanced with greater activity. It therefore cannot be renormalized through an effective temperature that implicitly treats the effect of the activity as enhanced unbiased random noise of the dimer particles.

Given the central nature of the WCA interaction, $F_{{\rm WCA},\bot}$ is necessarily antisymmetric in $y$. Therefore, the deflection associated with the odd mobility must be the result of an asymmetry in the PDF under an applied force for finite ${\rm Pe}$ such that the integral in Eq. \ref{eqn:orthogonal_velocity} is nonvanishing. Indeed, this is observed in Figs. \ref{fig:Fig3}C and D, where we plot the PDF in the upper half plane less that in the lower half plane, $g(x,y) - g(x,-y)$. This is a direct measure of the asymmetry under the transformation $y \rightarrow - y$ and hence of a nonvanishing odd mobility. We observe that there is measurable asymmetry only in the case of finite P\'{e}clet number, and the odd mobility is therefore due to asymmetric deformations of the particle distribution about the tracer particle for finite activities.

In this letter, we have derived a generalized Green-Kubo relation relating current and frenesy fluctuations in a steady state far from equilibrium to the mobility of a passive tracer suspended in a two-dimensional chiral active fluid. We have validated our results using molecular dynamics simulations and shown that, whereas the even mobility counterintuitively decreases with increasing activity, the odd mobility increases until it saturates to a finite value. The former result indicates that an effective temperature description of the mobility response is invalid in this system. We have further shown that the odd mobility decays more rapidly than the even mobility with increasing tracer particle size, independent of activity, indicating the possibility of size-based particle separation using an odd viscous working fluid. Our results provide strong evidence for the generality of the path integral framework for nonequilibrium response and a microscopic picture of the mobility response. Future work will focus on the development of effective hydrodynamic descriptions of the underlying fluid and tracer mobility.

\vspace{0.5cm}
\noindent{\bf Acknowledgments}
A. R. P. and D. T. L. acknowledge helpful discussions with Kranthi Mandadapu and Cory Hargus. This work has been supported by NSF Grant CHE-1954580. A. R. P  was also supported by the Heising-Simons Fellowship from the Kavli Energy Nanoscience Institute at UC Berkeley and D. T. L acknowledges support from thr Alfred P. Sloan Foundation.
\vspace{0.5cm}


\bibliography{mobility_ref}

\begin{thebibliography}{35}%
\makeatletter
\providecommand \@ifxundefined [1]{%
 \@ifx{#1\undefined}
}%
\providecommand \@ifnum [1]{%
 \ifnum #1\expandafter \@firstoftwo
 \else \expandafter \@secondoftwo
 \fi
}%
\providecommand \@ifx [1]{%
 \ifx #1\expandafter \@firstoftwo
 \else \expandafter \@secondoftwo
 \fi
}%
\providecommand \natexlab [1]{#1}%
\providecommand \enquote  [1]{``#1''}%
\providecommand \bibnamefont  [1]{#1}%
\providecommand \bibfnamefont [1]{#1}%
\providecommand \citenamefont [1]{#1}%
\providecommand \href@noop [0]{\@secondoftwo}%
\providecommand \href [0]{\begingroup \@sanitize@url \@href}%
\providecommand \@href[1]{\@@startlink{#1}\@@href}%
\providecommand \@@href[1]{\endgroup#1\@@endlink}%
\providecommand \@sanitize@url [0]{\catcode `\\12\catcode `\$12\catcode
  `\&12\catcode `\#12\catcode `\^12\catcode `\_12\catcode `\%12\relax}%
\providecommand \@@startlink[1]{}%
\providecommand \@@endlink[0]{}%
\providecommand \url  [0]{\begingroup\@sanitize@url \@url }%
\providecommand \@url [1]{\endgroup\@href {#1}{\urlprefix }}%
\providecommand \urlprefix  [0]{URL }%
\providecommand \Eprint [0]{\href }%
\providecommand \doibase [0]{http://dx.doi.org/}%
\providecommand \selectlanguage [0]{\@gobble}%
\providecommand \bibinfo  [0]{\@secondoftwo}%
\providecommand \bibfield  [0]{\@secondoftwo}%
\providecommand \translation [1]{[#1]}%
\providecommand \BibitemOpen [0]{}%
\providecommand \bibitemStop [0]{}%
\providecommand \bibitemNoStop [0]{.\EOS\space}%
\providecommand \EOS [0]{\spacefactor3000\relax}%
\providecommand \BibitemShut  [1]{\csname bibitem#1\endcsname}%
\let\auto@bib@innerbib\@empty
\bibitem [{\citenamefont {Banerjee}\ \emph {et~al.}(2017)\citenamefont
  {Banerjee}, \citenamefont {Souslov}, \citenamefont {Abanov},\ and\
  \citenamefont {Vitelli}}]{Banerjee}%
  \BibitemOpen
  \bibfield  {author} {\bibinfo {author} {\bibfnamefont {D.}~\bibnamefont
  {Banerjee}}, \bibinfo {author} {\bibfnamefont {A.}~\bibnamefont {Souslov}},
  \bibinfo {author} {\bibfnamefont {A.~G.}\ \bibnamefont {Abanov}}, \ and\
  \bibinfo {author} {\bibfnamefont {V.}~\bibnamefont {Vitelli}},\ }\bibfield
  {title} {\enquote {\bibinfo {title} {Odd viscosity in chiral active
  fluids},}\ }\href {\doibase 10.1038/s41467-017-01378-7} {\bibfield  {journal}
  {\bibinfo  {journal} {Nat. Comm.}\ }\textbf {\bibinfo {volume} {8}},\
  \bibinfo {pages} {1573} (\bibinfo {year} {2017})}\BibitemShut {NoStop}%
\bibitem [{\citenamefont {Hargus}\ \emph {et~al.}(2020)\citenamefont {Hargus},
  \citenamefont {Klymko}, \citenamefont {Epstein},\ and\ \citenamefont
  {Mandadapu}}]{Cory_MD}%
  \BibitemOpen
  \bibfield  {author} {\bibinfo {author} {\bibfnamefont {C.}~\bibnamefont
  {Hargus}}, \bibinfo {author} {\bibfnamefont {K.}~\bibnamefont {Klymko}},
  \bibinfo {author} {\bibfnamefont {J.~M.}\ \bibnamefont {Epstein}}, \ and\
  \bibinfo {author} {\bibfnamefont {K.~K.}\ \bibnamefont {Mandadapu}},\
  }\bibfield  {title} {\enquote {\bibinfo {title} {Time reversal symmetry
  breaking and odd viscosity in active fluids: {G}reen--{K}ubo and {NEMD}
  results},}\ }\href {\doibase 10.1063/5.0006441} {\bibfield  {journal}
  {\bibinfo  {journal} {J. Chem. Phys.}\ }\textbf {\bibinfo {volume} {152}},\
  \bibinfo {pages} {201102} (\bibinfo {year} {2020})}\BibitemShut {NoStop}%
\bibitem [{\citenamefont {Han}\ \emph {et~al.}(2021)\citenamefont {Han},
  \citenamefont {Fruchart}, \citenamefont {Scheibner}, \citenamefont
  {Vaikuntanathan}, \citenamefont {{d}e Pablo},\ and\ \citenamefont
  {Vitelli}}]{Vitellis_pretty_pictures}%
  \BibitemOpen
  \bibfield  {author} {\bibinfo {author} {\bibfnamefont {M.}~\bibnamefont
  {Han}}, \bibinfo {author} {\bibfnamefont {M.}~\bibnamefont {Fruchart}},
  \bibinfo {author} {\bibfnamefont {C.}~\bibnamefont {Scheibner}}, \bibinfo
  {author} {\bibfnamefont {S.}~\bibnamefont {Vaikuntanathan}}, \bibinfo
  {author} {\bibfnamefont {J.~J.}\ \bibnamefont {{d}e Pablo}}, \ and\ \bibinfo
  {author} {\bibfnamefont {V.}~\bibnamefont {Vitelli}},\ }\bibfield  {title}
  {\enquote {\bibinfo {title} {Fluctuating hydrodynamics of chiral active
  fluids},}\ }\href {\doibase 10.1038/s41567-021-01360-7} {\bibfield  {journal}
  {\bibinfo  {journal} {Nat. Phys.}\ }\textbf {\bibinfo {volume} {17}},\
  \bibinfo {pages} {1260--1269} (\bibinfo {year} {2021})}\BibitemShut {NoStop}%
\bibitem [{\citenamefont {Avron}(1998)}]{Avron}%
  \BibitemOpen
  \bibfield  {author} {\bibinfo {author} {\bibfnamefont {J.~E.}\ \bibnamefont
  {Avron}},\ }\bibfield  {title} {\enquote {\bibinfo {title} {Odd viscosity},}\
  }\href {\doibase 10.1023/A:1023084404080} {\bibfield  {journal} {\bibinfo
  {journal} {J. Stat. Phys.}\ }\textbf {\bibinfo {volume} {92}},\ \bibinfo
  {pages} {543--557} (\bibinfo {year} {1998})}\BibitemShut {NoStop}%
\bibitem [{\citenamefont {Dasbiswas}\ \emph {et~al.}(2019)\citenamefont
  {Dasbiswas}, \citenamefont {Mandadapu},\ and\ \citenamefont
  {Vaikuntanathan}}]{waves_Kranthi}%
  \BibitemOpen
  \bibfield  {author} {\bibinfo {author} {\bibfnamefont {K.}~\bibnamefont
  {Dasbiswas}}, \bibinfo {author} {\bibfnamefont {K.~K.}\ \bibnamefont
  {Mandadapu}}, \ and\ \bibinfo {author} {\bibfnamefont {S.}~\bibnamefont
  {Vaikuntanathan}},\ }\bibfield  {title} {\enquote {\bibinfo {title}
  {Topological localization in out-of-equilibrium dissipative systems},}\
  }\href {\doibase 10.1073/pnas.1721096115} {\bibfield  {journal} {\bibinfo
  {journal} {Proc. Natl. Acad. Sci. U.S.A.}\ }\textbf {\bibinfo {volume}
  {115}},\ \bibinfo {pages} {{E}9031 -- {E}9040} (\bibinfo {year}
  {2019})}\BibitemShut {NoStop}%
\bibitem [{\citenamefont {Ganeshan}\ and\ \citenamefont
  {Abanov}(2017)}]{bubble_dynamics_odd_viscosity}%
  \BibitemOpen
  \bibfield  {author} {\bibinfo {author} {\bibfnamefont {S.}~\bibnamefont
  {Ganeshan}}\ and\ \bibinfo {author} {\bibfnamefont {A.~G.}\ \bibnamefont
  {Abanov}},\ }\bibfield  {title} {\enquote {\bibinfo {title} {Odd viscosity in
  two-dimensional incompressible fluids},}\ }\href {\doibase
  10.1103/PhysRevFluids.2.094101} {\bibfield  {journal} {\bibinfo  {journal}
  {Phys. Rev. Fluids}\ }\textbf {\bibinfo {volume} {2}},\ \bibinfo {pages}
  {094101} (\bibinfo {year} {2017})}\BibitemShut {NoStop}%
\bibitem [{\citenamefont {Hargus}\ \emph {et~al.}(2021)\citenamefont {Hargus},
  \citenamefont {Epstein},\ and\ \citenamefont
  {Mandadapu}}]{Cory_odd_diffusivity}%
  \BibitemOpen
  \bibfield  {author} {\bibinfo {author} {\bibfnamefont {C.}~\bibnamefont
  {Hargus}}, \bibinfo {author} {\bibfnamefont {J.~M.}\ \bibnamefont {Epstein}},
  \ and\ \bibinfo {author} {\bibfnamefont {K.~K.}\ \bibnamefont {Mandadapu}},\
  }\bibfield  {title} {\enquote {\bibinfo {title} {Odd diffusivity of chiral
  random motion},}\ }\href {\doibase 10.1103/PhysRevLett.127.178001} {\bibfield
   {journal} {\bibinfo  {journal} {Phys. Rev. Lett.}\ }\textbf {\bibinfo
  {volume} {127}},\ \bibinfo {pages} {178001} (\bibinfo {year}
  {2021})}\BibitemShut {NoStop}%
\bibitem [{\citenamefont {Hosaka}\ \emph {et~al.}(2021)\citenamefont {Hosaka},
  \citenamefont {Komura},\ and\ \citenamefont {Andelman}}]{Hosaka_mobility}%
  \BibitemOpen
  \bibfield  {author} {\bibinfo {author} {\bibfnamefont {Y.}~\bibnamefont
  {Hosaka}}, \bibinfo {author} {\bibfnamefont {S.}~\bibnamefont {Komura}}, \
  and\ \bibinfo {author} {\bibfnamefont {D.}~\bibnamefont {Andelman}},\
  }\bibfield  {title} {\enquote {\bibinfo {title} {Nonreciprocal response of a
  two-dimensional fluid with odd viscosity},}\ }\href {\doibase
  10.1103/PhysRevE.103.042610} {\bibfield  {journal} {\bibinfo  {journal}
  {Phys. Rev. E}\ }\textbf {\bibinfo {volume} {103}},\ \bibinfo {pages}
  {042610} (\bibinfo {year} {2021})}\BibitemShut {NoStop}%
\bibitem [{\citenamefont {Lou}\ \emph {et~al.}(2022{\natexlab{a}})\citenamefont
  {Lou}, \citenamefont {Yang}, \citenamefont {Ding}, \citenamefont {Liu},
  \citenamefont {Chen}, \citenamefont {Zhou}, \citenamefont {Ye}, \citenamefont
  {Podgornik},\ and\ \citenamefont {Yang}}]{orthogonal_fluid_motion}%
  \BibitemOpen
  \bibfield  {author} {\bibinfo {author} {\bibfnamefont {X.}~\bibnamefont
  {Lou}}, \bibinfo {author} {\bibfnamefont {Q.}~\bibnamefont {Yang}}, \bibinfo
  {author} {\bibfnamefont {Y.}~\bibnamefont {Ding}}, \bibinfo {author}
  {\bibfnamefont {P.}~\bibnamefont {Liu}}, \bibinfo {author} {\bibfnamefont
  {K.}~\bibnamefont {Chen}}, \bibinfo {author} {\bibfnamefont {X.}~\bibnamefont
  {Zhou}}, \bibinfo {author} {\bibfnamefont {F.}~\bibnamefont {Ye}}, \bibinfo
  {author} {\bibfnamefont {R.}~\bibnamefont {Podgornik}}, \ and\ \bibinfo
  {author} {\bibfnamefont {M.}~\bibnamefont {Yang}},\ }\bibfield  {title}
  {\enquote {\bibinfo {title} {Odd viscosity-induced hall-like transport of an
  active chiral fluid},}\ }\href {\doibase 10.1073/pnas.2201279119} {\bibfield
  {journal} {\bibinfo  {journal} {Proc. Natl. Acad. Sci. U.S.A.}\ }\textbf
  {\bibinfo {volume} {119}},\ \bibinfo {pages} {e2201279119} (\bibinfo {year}
  {2022}{\natexlab{a}})}\BibitemShut {NoStop}%
\bibitem [{\citenamefont {Markovich}\ and\ \citenamefont
  {Lubensky}(2021)}]{3d_odd_viscosity}%
  \BibitemOpen
  \bibfield  {author} {\bibinfo {author} {\bibfnamefont {T.}~\bibnamefont
  {Markovich}}\ and\ \bibinfo {author} {\bibfnamefont {T.~C.}\ \bibnamefont
  {Lubensky}},\ }\bibfield  {title} {\enquote {\bibinfo {title} {Odd viscosity
  in active matter: {M}icroscopic origin and 3d effects},}\ }\href {\doibase
  10.1103/PhysRevLett.127.048001} {\bibfield  {journal} {\bibinfo  {journal}
  {Phys. Rev. Lett.}\ }\textbf {\bibinfo {volume} {127}},\ \bibinfo {pages}
  {048001} (\bibinfo {year} {2021})}\BibitemShut {NoStop}%
\bibitem [{\citenamefont {Souslov}\ \emph {et~al.}(2019)\citenamefont
  {Souslov}, \citenamefont {Dasbiswas}, \citenamefont {Fruchart}, \citenamefont
  {Vaikuntanathan},\ and\ \citenamefont {Vitelli}}]{boundary_waves}%
  \BibitemOpen
  \bibfield  {author} {\bibinfo {author} {\bibfnamefont {A.}~\bibnamefont
  {Souslov}}, \bibinfo {author} {\bibfnamefont {K.}~\bibnamefont {Dasbiswas}},
  \bibinfo {author} {\bibfnamefont {M.}~\bibnamefont {Fruchart}}, \bibinfo
  {author} {\bibfnamefont {S.}~\bibnamefont {Vaikuntanathan}}, \ and\ \bibinfo
  {author} {\bibfnamefont {V.}~\bibnamefont {Vitelli}},\ }\bibfield  {title}
  {\enquote {\bibinfo {title} {Topological waves in fluids with odd
  viscosity},}\ }\href {\doibase 10.1103/PhysRevLett.122.128001} {\bibfield
  {journal} {\bibinfo  {journal} {Phys. Rev. Lett.}\ }\textbf {\bibinfo
  {volume} {122}},\ \bibinfo {pages} {128001} (\bibinfo {year}
  {2019})}\BibitemShut {NoStop}%
\bibitem [{\citenamefont {{v}an Zuiden}\ \emph {et~al.}(2016)\citenamefont
  {{v}an Zuiden}, \citenamefont {Paulose}, \citenamefont {Irvine},
  \citenamefont {Bartolo},\ and\ \citenamefont {Vitelli}}]{crystal}%
  \BibitemOpen
  \bibfield  {author} {\bibinfo {author} {\bibfnamefont {B.~C.}\ \bibnamefont
  {{v}an Zuiden}}, \bibinfo {author} {\bibfnamefont {J.}~\bibnamefont
  {Paulose}}, \bibinfo {author} {\bibfnamefont {W.~T.~M.}\ \bibnamefont
  {Irvine}}, \bibinfo {author} {\bibfnamefont {D.}~\bibnamefont {Bartolo}}, \
  and\ \bibinfo {author} {\bibfnamefont {V.}~\bibnamefont {Vitelli}},\
  }\bibfield  {title} {\enquote {\bibinfo {title} {Spatiotemporal order and
  emergent edge currents in active spinner materials},}\ }\href {\doibase
  10.1073/pnas.1609572113} {\bibfield  {journal} {\bibinfo  {journal} {Proc.
  Natl. Acad. Sci. U.S.A.}\ }\textbf {\bibinfo {volume} {113}},\ \bibinfo
  {pages} {12919--12924} (\bibinfo {year} {2016})}\BibitemShut {NoStop}%
\bibitem [{\citenamefont {Yang}\ \emph {et~al.}(2021)\citenamefont {Yang},
  \citenamefont {Zhu}, \citenamefont {Liu}, \citenamefont {Liu}, \citenamefont
  {Shi}, \citenamefont {Chen}, \citenamefont {Zheng}, \citenamefont {Ye},\ and\
  \citenamefont {Yang}}]{cargo}%
  \BibitemOpen
  \bibfield  {author} {\bibinfo {author} {\bibfnamefont {Q.}~\bibnamefont
  {Yang}}, \bibinfo {author} {\bibfnamefont {H.}~\bibnamefont {Zhu}}, \bibinfo
  {author} {\bibfnamefont {P.}~\bibnamefont {Liu}}, \bibinfo {author}
  {\bibfnamefont {R.}~\bibnamefont {Liu}}, \bibinfo {author} {\bibfnamefont
  {Q.}~\bibnamefont {Shi}}, \bibinfo {author} {\bibfnamefont {K.}~\bibnamefont
  {Chen}}, \bibinfo {author} {\bibfnamefont {N.}~\bibnamefont {Zheng}},
  \bibinfo {author} {\bibfnamefont {F.}~\bibnamefont {Ye}}, \ and\ \bibinfo
  {author} {\bibfnamefont {M.}~\bibnamefont {Yang}},\ }\bibfield  {title}
  {\enquote {\bibinfo {title} {Topologically protected transport of cargo in a
  chiral active fluid aided by odd-viscosity-enhanced depletion
  interactions},}\ }\href {\doibase 10.1103/PhysRevLett.126.198001} {\bibfield
  {journal} {\bibinfo  {journal} {Phys. Rev. Lett.}\ }\textbf {\bibinfo
  {volume} {126}},\ \bibinfo {pages} {198001} (\bibinfo {year}
  {2021})}\BibitemShut {NoStop}%
\bibitem [{\citenamefont {Lou}\ \emph {et~al.}(2022{\natexlab{b}})\citenamefont
  {Lou}, \citenamefont {Yang}, \citenamefont {Ding}, \citenamefont {Liu},
  \citenamefont {Chen}, \citenamefont {Zhou}, \citenamefont {Ye}, \citenamefont
  {Podgornik},\ and\ \citenamefont {Yang}}]{Hall-like}%
  \BibitemOpen
  \bibfield  {author} {\bibinfo {author} {\bibfnamefont {X.}~\bibnamefont
  {Lou}}, \bibinfo {author} {\bibfnamefont {Q.}~\bibnamefont {Yang}}, \bibinfo
  {author} {\bibfnamefont {Y.}~\bibnamefont {Ding}}, \bibinfo {author}
  {\bibfnamefont {P.}~\bibnamefont {Liu}}, \bibinfo {author} {\bibfnamefont
  {K.}~\bibnamefont {Chen}}, \bibinfo {author} {\bibfnamefont {X.}~\bibnamefont
  {Zhou}}, \bibinfo {author} {\bibfnamefont {F.}~\bibnamefont {Ye}}, \bibinfo
  {author} {\bibfnamefont {R.}~\bibnamefont {Podgornik}}, \ and\ \bibinfo
  {author} {\bibfnamefont {M.}~\bibnamefont {Yang}},\ }\bibfield  {title}
  {\enquote {\bibinfo {title} {Odd viscosity-induced hall-like transport of an
  active chiral fluid},}\ }\href {\doibase 10.1073/pnas.2201279119} {\bibfield
  {journal} {\bibinfo  {journal} {Proc. Natl. Acad. Sci. U.S.A.}\ }\textbf
  {\bibinfo {volume} {119}},\ \bibinfo {pages} {e2201279119} (\bibinfo {year}
  {2022}{\natexlab{b}})}\BibitemShut {NoStop}%
\bibitem [{\citenamefont {Soni}\ \emph {et~al.}(2019)\citenamefont {Soni},
  \citenamefont {Bililign}, \citenamefont {Magkiriadou}, \citenamefont
  {Sacanna}, \citenamefont {Bartolo}, \citenamefont {Shelley},\ and\
  \citenamefont {Irvine}}]{free_surface}%
  \BibitemOpen
  \bibfield  {author} {\bibinfo {author} {\bibfnamefont {V.}~\bibnamefont
  {Soni}}, \bibinfo {author} {\bibfnamefont {E.~S.}\ \bibnamefont {Bililign}},
  \bibinfo {author} {\bibfnamefont {S.}~\bibnamefont {Magkiriadou}}, \bibinfo
  {author} {\bibfnamefont {S.}~\bibnamefont {Sacanna}}, \bibinfo {author}
  {\bibfnamefont {D.}~\bibnamefont {Bartolo}}, \bibinfo {author} {\bibfnamefont
  {M.~J.}\ \bibnamefont {Shelley}}, \ and\ \bibinfo {author} {\bibfnamefont
  {W.~T.~M.}\ \bibnamefont {Irvine}},\ }\bibfield  {title} {\enquote {\bibinfo
  {title} {The odd free surface flows of a colloidal chiral fluid},}\ }\href
  {\doibase 10.1038/s41567-019-0603-8} {\bibfield  {journal} {\bibinfo
  {journal} {Nat. Phys.}\ }\textbf {\bibinfo {volume} {15}},\ \bibinfo {pages}
  {1188--1194} (\bibinfo {year} {2019})}\BibitemShut {NoStop}%
\bibitem [{\citenamefont {De{G}root}\ and\ \citenamefont
  {Mazur}(2011)}]{deGroot+Mazur}%
  \BibitemOpen
  \bibfield  {author} {\bibinfo {author} {\bibfnamefont {S.~R.}\ \bibnamefont
  {De{G}root}}\ and\ \bibinfo {author} {\bibfnamefont {P.}~\bibnamefont
  {Mazur}},\ }\href@noop {} {\emph {\bibinfo {title} {Non-Equilibrium
  Thermodynamics}}},\ \bibinfo {edition} {2nd}\ ed.\ (\bibinfo  {publisher}
  {Dover},\ \bibinfo {year} {2011})\ p.\ \bibinfo {pages} {528}\BibitemShut
  {NoStop}%
\bibitem [{\citenamefont {Gao}\ and\ \citenamefont {Limmer}(2017)}]{Chloe1}%
  \BibitemOpen
  \bibfield  {author} {\bibinfo {author} {\bibfnamefont {C.~Y.}\ \bibnamefont
  {Gao}}\ and\ \bibinfo {author} {\bibfnamefont {D.~T.}\ \bibnamefont
  {Limmer}},\ }\bibfield  {title} {\enquote {\bibinfo {title} {Transport
  coefficients from large deviation functions},}\ }\href {\doibase
  10.3390/e19110571} {\bibfield  {journal} {\bibinfo  {journal} {Entropy}\
  }\textbf {\bibinfo {volume} {19}} (\bibinfo {year} {2017}),\
  10.3390/e19110571}\BibitemShut {NoStop}%
\bibitem [{\citenamefont {Gao}\ and\ \citenamefont {Limmer}(2019)}]{Chloe2}%
  \BibitemOpen
  \bibfield  {author} {\bibinfo {author} {\bibfnamefont {C.~Y.}\ \bibnamefont
  {Gao}}\ and\ \bibinfo {author} {\bibfnamefont {D.~T.}\ \bibnamefont
  {Limmer}},\ }\bibfield  {title} {\enquote {\bibinfo {title} {Nonlinear
  transport coefficients from large deviation functions},}\ }\href {\doibase
  10.1063/1.5110507} {\bibfield  {journal} {\bibinfo  {journal} {J. Chem.
  Phys.}\ }\textbf {\bibinfo {volume} {151}},\ \bibinfo {pages} {014101}
  (\bibinfo {year} {2019})}\BibitemShut {NoStop}%
\bibitem [{\citenamefont {Limmer}\ \emph {et~al.}(2021)\citenamefont {Limmer},
  \citenamefont {Gao},\ and\ \citenamefont {Poggioli}}]{review}%
  \BibitemOpen
  \bibfield  {author} {\bibinfo {author} {\bibfnamefont {D.~T.}\ \bibnamefont
  {Limmer}}, \bibinfo {author} {\bibfnamefont {C.~Y.}\ \bibnamefont {Gao}}, \
  and\ \bibinfo {author} {\bibfnamefont {A.~R.}\ \bibnamefont {Poggioli}},\
  }\bibfield  {title} {\enquote {\bibinfo {title} {A large deviation theory
  perspective on nanoscale transport phenomena},}\ }\href {\doibase
  10.1140/epjb/s10051-021-00164-1} {\bibfield  {journal} {\bibinfo  {journal}
  {Eur. Phys. J. B}\ }\textbf {\bibinfo {volume} {94}},\ \bibinfo {pages} {145}
  (\bibinfo {year} {2021})}\BibitemShut {NoStop}%
\bibitem [{\citenamefont {Weeks}\ \emph {et~al.}(1971)\citenamefont {Weeks},
  \citenamefont {Chandler},\ and\ \citenamefont {Andersen}}]{WCA}%
  \BibitemOpen
  \bibfield  {author} {\bibinfo {author} {\bibfnamefont {J.~D.}\ \bibnamefont
  {Weeks}}, \bibinfo {author} {\bibfnamefont {D.}~\bibnamefont {Chandler}}, \
  and\ \bibinfo {author} {\bibfnamefont {H.~C.}\ \bibnamefont {Andersen}},\
  }\bibfield  {title} {\enquote {\bibinfo {title} {Role of repulsive forces in
  determining the equilibrium structure of simple liquids},}\ }\href {\doibase
  10.1063/1.1674820} {\bibfield  {journal} {\bibinfo  {journal} {J. Chem.
  Phys.}\ }\textbf {\bibinfo {volume} {54}},\ \bibinfo {pages} {5237--5247}
  (\bibinfo {year} {1971})}\BibitemShut {NoStop}%
\bibitem [{\citenamefont {Marchetti}\ \emph {et~al.}(2013)\citenamefont
  {Marchetti}, \citenamefont {Joanny}, \citenamefont {Ramaswamy}, \citenamefont
  {Liverpool}, \citenamefont {Prost}, \citenamefont {Rao},\ and\ \citenamefont
  {Simha}}]{joanny}%
  \BibitemOpen
  \bibfield  {author} {\bibinfo {author} {\bibfnamefont {M.~C.}\ \bibnamefont
  {Marchetti}}, \bibinfo {author} {\bibfnamefont {J.~F.}\ \bibnamefont
  {Joanny}}, \bibinfo {author} {\bibfnamefont {S.}~\bibnamefont {Ramaswamy}},
  \bibinfo {author} {\bibfnamefont {T.~B.}\ \bibnamefont {Liverpool}}, \bibinfo
  {author} {\bibfnamefont {J.}~\bibnamefont {Prost}}, \bibinfo {author}
  {\bibfnamefont {Madan}\ \bibnamefont {Rao}}, \ and\ \bibinfo {author}
  {\bibfnamefont {R.~Aditi}\ \bibnamefont {Simha}},\ }\bibfield  {title}
  {\enquote {\bibinfo {title} {Hydrodynamics of soft active matter},}\ }\href
  {\doibase 10.1103/RevModPhys.85.1143} {\bibfield  {journal} {\bibinfo
  {journal} {Rev. Mod. Phys.}\ }\textbf {\bibinfo {volume} {85}},\ \bibinfo
  {pages} {1143--1189} (\bibinfo {year} {2013})}\BibitemShut {NoStop}%
\bibitem [{\citenamefont {Weisenborn}\ and\ \citenamefont
  {Mazur}(1984)}]{2d_Stokes}%
  \BibitemOpen
  \bibfield  {author} {\bibinfo {author} {\bibfnamefont {A.~J.}\ \bibnamefont
  {Weisenborn}}\ and\ \bibinfo {author} {\bibfnamefont {P.}~\bibnamefont
  {Mazur}},\ }\bibfield  {title} {\enquote {\bibinfo {title} {The {O}seen drag
  on a circular cylinder revisited},}\ }\href {\doibase
  10.1016/0378-4371(84)90111-0} {\bibfield  {journal} {\bibinfo  {journal}
  {Physica A}\ }\textbf {\bibinfo {volume} {123}},\ \bibinfo {pages} {191--208}
  (\bibinfo {year} {1984})}\BibitemShut {NoStop}%
\bibitem [{\citenamefont {Hargus}\ and\ \citenamefont
  {Mandadapu}(2022)}]{Cory_github}%
  \BibitemOpen
  \bibfield  {author} {\bibinfo {author} {\bibfnamefont {C.}~\bibnamefont
  {Hargus}}\ and\ \bibinfo {author} {\bibfnamefont {K.~K.}\ \bibnamefont
  {Mandadapu}},\ }\href@noop {} {\enquote {\bibinfo {title} {Molecular
  simulations of active matter},}\ }\bibinfo {howpublished}
  {https://github.com/mandadapu-group/active-matter} (\bibinfo {year}
  {2022})\BibitemShut {NoStop}%
\bibitem [{\citenamefont {Cugliandolo}\ \emph {et~al.}(2019)\citenamefont
  {Cugliandolo}, \citenamefont {Lecomte},\ and\ \citenamefont
  {Van~Wijland}}]{cugliandolo2019building}%
  \BibitemOpen
  \bibfield  {author} {\bibinfo {author} {\bibfnamefont {Leticia~F}\
  \bibnamefont {Cugliandolo}}, \bibinfo {author} {\bibfnamefont {Vivien}\
  \bibnamefont {Lecomte}}, \ and\ \bibinfo {author} {\bibfnamefont
  {Fr{\'e}d{\'e}ric}\ \bibnamefont {Van~Wijland}},\ }\bibfield  {title}
  {\enquote {\bibinfo {title} {Building a path-integral calculus: a covariant
  discretization approach},}\ }\href@noop {} {\bibfield  {journal} {\bibinfo
  {journal} {Journal of Physics A: Mathematical and Theoretical}\ }\textbf
  {\bibinfo {volume} {52}},\ \bibinfo {pages} {50LT01} (\bibinfo {year}
  {2019})}\BibitemShut {NoStop}%
\bibitem [{\citenamefont {Onsager}\ and\ \citenamefont
  {Machlup}(1953)}]{Onsager+Machlup1}%
  \BibitemOpen
  \bibfield  {author} {\bibinfo {author} {\bibfnamefont {L.}~\bibnamefont
  {Onsager}}\ and\ \bibinfo {author} {\bibfnamefont {S.}~\bibnamefont
  {Machlup}},\ }\bibfield  {title} {\enquote {\bibinfo {title} {Fluctuations
  and irreversible processes},}\ }\href {\doibase 10.1103/PhysRev.91.1505}
  {\bibfield  {journal} {\bibinfo  {journal} {Phys. Rev.}\ }\textbf {\bibinfo
  {volume} {91}},\ \bibinfo {pages} {1505--1512} (\bibinfo {year}
  {1953})}\BibitemShut {NoStop}%
\bibitem [{\citenamefont {Machlup}\ and\ \citenamefont
  {Onsager}(1953)}]{Onsager+Machlup2}%
  \BibitemOpen
  \bibfield  {author} {\bibinfo {author} {\bibfnamefont {S.}~\bibnamefont
  {Machlup}}\ and\ \bibinfo {author} {\bibfnamefont {L.}~\bibnamefont
  {Onsager}},\ }\bibfield  {title} {\enquote {\bibinfo {title} {Fluctuations
  and irreversible process. ii. systems with kinetic energy},}\ }\href
  {\doibase 10.1103/PhysRev.91.1512} {\bibfield  {journal} {\bibinfo  {journal}
  {Phys. Rev.}\ }\textbf {\bibinfo {volume} {91}},\ \bibinfo {pages}
  {1512--1515} (\bibinfo {year} {1953})}\BibitemShut {NoStop}%
\bibitem [{\citenamefont {Basu}\ and\ \citenamefont
  {Maes}(2015)}]{basu2015nonequilibrium}%
  \BibitemOpen
  \bibfield  {author} {\bibinfo {author} {\bibfnamefont {Urna}\ \bibnamefont
  {Basu}}\ and\ \bibinfo {author} {\bibfnamefont {Christian}\ \bibnamefont
  {Maes}},\ }\bibfield  {title} {\enquote {\bibinfo {title} {Nonequilibrium
  response and frenesy},}\ }in\ \href@noop {} {\emph {\bibinfo {booktitle}
  {Journal of Physics: Conference Series}}},\ Vol.\ \bibinfo {volume} {638}\
  (\bibinfo {organization} {IOP Publishing},\ \bibinfo {year} {2015})\ p.\
  \bibinfo {pages} {012001}\BibitemShut {NoStop}%
\bibitem [{\citenamefont {Baiesi}\ \emph {et~al.}(2009)\citenamefont {Baiesi},
  \citenamefont {Maes},\ and\ \citenamefont {Wynants}}]{Maes_frenesy}%
  \BibitemOpen
  \bibfield  {author} {\bibinfo {author} {\bibfnamefont {Marco}\ \bibnamefont
  {Baiesi}}, \bibinfo {author} {\bibfnamefont {Christian}\ \bibnamefont
  {Maes}}, \ and\ \bibinfo {author} {\bibfnamefont {Bram}\ \bibnamefont
  {Wynants}},\ }\bibfield  {title} {\enquote {\bibinfo {title} {Fluctuations
  and response of nonequilibrium states},}\ }\href {\doibase
  10.1103/PhysRevLett.103.010602} {\bibfield  {journal} {\bibinfo  {journal}
  {Phys. Rev. Lett.}\ }\textbf {\bibinfo {volume} {103}},\ \bibinfo {pages}
  {010602} (\bibinfo {year} {2009})}\BibitemShut {NoStop}%
\bibitem [{\citenamefont {Lesnicki}\ \emph {et~al.}(2020)\citenamefont
  {Lesnicki}, \citenamefont {Gao}, \citenamefont {Rotenberg},\ and\
  \citenamefont {Limmer}}]{Chloe3_PDF}%
  \BibitemOpen
  \bibfield  {author} {\bibinfo {author} {\bibfnamefont {D.}~\bibnamefont
  {Lesnicki}}, \bibinfo {author} {\bibfnamefont {C.~Y.}\ \bibnamefont {Gao}},
  \bibinfo {author} {\bibfnamefont {B.}~\bibnamefont {Rotenberg}}, \ and\
  \bibinfo {author} {\bibfnamefont {D.~T.}\ \bibnamefont {Limmer}},\ }\bibfield
   {title} {\enquote {\bibinfo {title} {Field-dependent ionic conductivities
  from generalized fluctuation-dissipation relations},}\ }\href {\doibase
  10.1103/PhysRevLett.124.206001} {\bibfield  {journal} {\bibinfo  {journal}
  {Phys. Rev. Lett.}\ }\textbf {\bibinfo {volume} {124}},\ \bibinfo {pages}
  {206001} (\bibinfo {year} {2020})}\BibitemShut {NoStop}%
\bibitem [{\citenamefont {Lesnicki}\ \emph {et~al.}(2021)\citenamefont
  {Lesnicki}, \citenamefont {Gao}, \citenamefont {Limmer},\ and\ \citenamefont
  {Rotenberg}}]{Chloe4_PDF}%
  \BibitemOpen
  \bibfield  {author} {\bibinfo {author} {\bibfnamefont {D.}~\bibnamefont
  {Lesnicki}}, \bibinfo {author} {\bibfnamefont {C.~Y.}\ \bibnamefont {Gao}},
  \bibinfo {author} {\bibfnamefont {D.~T.}\ \bibnamefont {Limmer}}, \ and\
  \bibinfo {author} {\bibfnamefont {B.}~\bibnamefont {Rotenberg}},\ }\bibfield
  {title} {\enquote {\bibinfo {title} {On the molecular correlations that
  result in field-dependent conductivities in electrolyte solutions},}\ }\href
  {\doibase 10.1063/5.0052860} {\bibfield  {journal} {\bibinfo  {journal} {J.
  Chem. Phys.}\ }\textbf {\bibinfo {volume} {155}},\ \bibinfo {pages} {014507}
  (\bibinfo {year} {2021})}\BibitemShut {NoStop}%
\bibitem [{\citenamefont {Reichhardt}\ and\ \citenamefont
  {Reichhardt}(2019)}]{Reichhardt1}%
  \BibitemOpen
  \bibfield  {author} {\bibinfo {author} {\bibfnamefont {C.}~\bibnamefont
  {Reichhardt}}\ and\ \bibinfo {author} {\bibfnamefont {C.~J.~O.}\ \bibnamefont
  {Reichhardt}},\ }\bibfield  {title} {\enquote {\bibinfo {title} {Active
  microrheology, hall effect, and jamming in chiral fluids},}\ }\href {\doibase
  10.1103/PhysRevE.100.012604} {\bibfield  {journal} {\bibinfo  {journal}
  {Phys. Rev. E}\ }\textbf {\bibinfo {volume} {100}},\ \bibinfo {pages}
  {012604} (\bibinfo {year} {2019})}\BibitemShut {NoStop}%
\bibitem [{\citenamefont {Reichhardt}\ and\ \citenamefont
  {Reichhardt}(2022)}]{Reichhardt2}%
  \BibitemOpen
  \bibfield  {author} {\bibinfo {author} {\bibfnamefont {C.~J.~O.}\
  \bibnamefont {Reichhardt}}\ and\ \bibinfo {author} {\bibfnamefont
  {C.}~\bibnamefont {Reichhardt}},\ }\bibfield  {title} {\enquote {\bibinfo
  {title} {Active rheology in odd-viscosity systems},}\ }\href {\doibase
  10.1209/0295-5075/ac2adc} {\bibfield  {journal} {\bibinfo  {journal}
  {Europhys. Lett.}\ }\textbf {\bibinfo {volume} {137}},\ \bibinfo {pages}
  {66004} (\bibinfo {year} {2022})}\BibitemShut {NoStop}%
\bibitem [{\citenamefont {Liao}\ \emph {et~al.}(2019)\citenamefont {Liao},
  \citenamefont {Han}, \citenamefont {Fruchart}, \citenamefont {Vitelli},\ and\
  \citenamefont {Vaikuntanathan}}]{Liao_PDFmechanics}%
  \BibitemOpen
  \bibfield  {author} {\bibinfo {author} {\bibfnamefont {Z.}~\bibnamefont
  {Liao}}, \bibinfo {author} {\bibfnamefont {M.}~\bibnamefont {Han}}, \bibinfo
  {author} {\bibfnamefont {M.}~\bibnamefont {Fruchart}}, \bibinfo {author}
  {\bibfnamefont {V.}~\bibnamefont {Vitelli}}, \ and\ \bibinfo {author}
  {\bibfnamefont {S.}~\bibnamefont {Vaikuntanathan}},\ }\bibfield  {title}
  {\enquote {\bibinfo {title} {A mechanism for anomalous transport in chiral
  active liquids},}\ }\href {\doibase 10.1063/1.5126962} {\bibfield  {journal}
  {\bibinfo  {journal} {J. Chem. Phys.}\ }\textbf {\bibinfo {volume} {151}},\
  \bibinfo {pages} {194108} (\bibinfo {year} {2019})}\BibitemShut {NoStop}%
\bibitem [{\citenamefont {Hansen}\ and\ \citenamefont
  {McDonald}(2013)}]{hansen2013theory}%
  \BibitemOpen
  \bibfield  {author} {\bibinfo {author} {\bibfnamefont {Jean-Pierre}\
  \bibnamefont {Hansen}}\ and\ \bibinfo {author} {\bibfnamefont {Ian~Ranald}\
  \bibnamefont {McDonald}},\ }\href@noop {} {\emph {\bibinfo {title} {Theory of
  simple liquids: with applications to soft matter}}}\ (\bibinfo  {publisher}
  {Academic press},\ \bibinfo {year} {2013})\BibitemShut {NoStop}%
\bibitem [{\citenamefont {Onsager}(1927)}]{onsager1927report}%
  \BibitemOpen
  \bibfield  {author} {\bibinfo {author} {\bibfnamefont {L}~\bibnamefont
  {Onsager}},\ }\bibfield  {title} {\enquote {\bibinfo {title} {Report on a
  revision of the conductivity theory},}\ }\href@noop {} {\bibfield  {journal}
  {\bibinfo  {journal} {Transactions of the Faraday Society}\ }\textbf
  {\bibinfo {volume} {23}},\ \bibinfo {pages} {341--349} (\bibinfo {year}
  {1927})}\BibitemShut {NoStop}%
\end{thebibliography}%


\begin{thebibliography}{7}%
\makeatletter
\providecommand \@ifxundefined [1]{%
 \@ifx{#1\undefined}
}%
\providecommand \@ifnum [1]{%
 \ifnum #1\expandafter \@firstoftwo
 \else \expandafter \@secondoftwo
 \fi
}%
\providecommand \@ifx [1]{%
 \ifx #1\expandafter \@firstoftwo
 \else \expandafter \@secondoftwo
 \fi
}%
\providecommand \natexlab [1]{#1}%
\providecommand \enquote  [1]{``#1''}%
\providecommand \bibnamefont  [1]{#1}%
\providecommand \bibfnamefont [1]{#1}%
\providecommand \citenamefont [1]{#1}%
\providecommand \href@noop [0]{\@secondoftwo}%
\providecommand \href [0]{\begingroup \@sanitize@url \@href}%
\providecommand \@href[1]{\@@startlink{#1}\@@href}%
\providecommand \@@href[1]{\endgroup#1\@@endlink}%
\providecommand \@sanitize@url [0]{\catcode `\\12\catcode `\$12\catcode
  `\&12\catcode `\#12\catcode `\^12\catcode `\_12\catcode `\%12\relax}%
\providecommand \@@startlink[1]{}%
\providecommand \@@endlink[0]{}%
\providecommand \url  [0]{\begingroup\@sanitize@url \@url }%
\providecommand \@url [1]{\endgroup\@href {#1}{\urlprefix }}%
\providecommand \urlprefix  [0]{URL }%
\providecommand \Eprint [0]{\href }%
\providecommand \doibase [0]{http://dx.doi.org/}%
\providecommand \selectlanguage [0]{\@gobble}%
\providecommand \bibinfo  [0]{\@secondoftwo}%
\providecommand \bibfield  [0]{\@secondoftwo}%
\providecommand \translation [1]{[#1]}%
\providecommand \BibitemOpen [0]{}%
\providecommand \bibitemStop [0]{}%
\providecommand \bibitemNoStop [0]{.\EOS\space}%
\providecommand \EOS [0]{\spacefactor3000\relax}%
\providecommand \BibitemShut  [1]{\csname bibitem#1\endcsname}%
\let\auto@bib@innerbib\@empty
\bibitem [{\citenamefont {Marchetti}\ \emph {et~al.}(2013)\citenamefont
  {Marchetti}, \citenamefont {Joanny}, \citenamefont {Ramaswamy}, \citenamefont
  {Liverpool}, \citenamefont {Prost}, \citenamefont {Rao},\ and\ \citenamefont
  {Simha}}]{joanny}%
  \BibitemOpen
  \bibfield  {author} {\bibinfo {author} {\bibfnamefont {M.~C.}\ \bibnamefont
  {Marchetti}}, \bibinfo {author} {\bibfnamefont {J.~F.}\ \bibnamefont
  {Joanny}}, \bibinfo {author} {\bibfnamefont {S.}~\bibnamefont {Ramaswamy}},
  \bibinfo {author} {\bibfnamefont {T.~B.}\ \bibnamefont {Liverpool}}, \bibinfo
  {author} {\bibfnamefont {J.}~\bibnamefont {Prost}}, \bibinfo {author}
  {\bibfnamefont {Madan}\ \bibnamefont {Rao}}, \ and\ \bibinfo {author}
  {\bibfnamefont {R.~Aditi}\ \bibnamefont {Simha}},\ }\bibfield  {title}
  {\enquote {\bibinfo {title} {Hydrodynamics of soft active matter},}\ }\href
  {\doibase 10.1103/RevModPhys.85.1143} {\bibfield  {journal} {\bibinfo
  {journal} {Rev. Mod. Phys.}\ }\textbf {\bibinfo {volume} {85}},\ \bibinfo
  {pages} {1143--1189} (\bibinfo {year} {2013})}\BibitemShut {NoStop}%
\bibitem [{\citenamefont {Weeks}\ \emph {et~al.}(1971)\citenamefont {Weeks},
  \citenamefont {Chandler},\ and\ \citenamefont {Andersen}}]{WCA}%
  \BibitemOpen
  \bibfield  {author} {\bibinfo {author} {\bibfnamefont {J.~D.}\ \bibnamefont
  {Weeks}}, \bibinfo {author} {\bibfnamefont {D.}~\bibnamefont {Chandler}}, \
  and\ \bibinfo {author} {\bibfnamefont {H.~C.}\ \bibnamefont {Andersen}},\
  }\bibfield  {title} {\enquote {\bibinfo {title} {Role of repulsive forces in
  determining the equilibrium structure of simple liquids},}\ }\href {\doibase
  10.1063/1.1674820} {\bibfield  {journal} {\bibinfo  {journal} {J. Chem.
  Phys.}\ }\textbf {\bibinfo {volume} {54}},\ \bibinfo {pages} {5237--5247}
  (\bibinfo {year} {1971})}\BibitemShut {NoStop}%
\bibitem [{\citenamefont {Hargus}\ \emph {et~al.}(2020)\citenamefont {Hargus},
  \citenamefont {Klymko}, \citenamefont {Epstein},\ and\ \citenamefont
  {Mandadapu}}]{Cory_MD}%
  \BibitemOpen
  \bibfield  {author} {\bibinfo {author} {\bibfnamefont {C.}~\bibnamefont
  {Hargus}}, \bibinfo {author} {\bibfnamefont {K.}~\bibnamefont {Klymko}},
  \bibinfo {author} {\bibfnamefont {J.~M.}\ \bibnamefont {Epstein}}, \ and\
  \bibinfo {author} {\bibfnamefont {K.~K.}\ \bibnamefont {Mandadapu}},\
  }\bibfield  {title} {\enquote {\bibinfo {title} {Time reversal symmetry
  breaking and odd viscosity in active fluids: {G}reen--{K}ubo and {NEMD}
  results},}\ }\href {\doibase 10.1063/5.0006441} {\bibfield  {journal}
  {\bibinfo  {journal} {J. Chem. Phys.}\ }\textbf {\bibinfo {volume} {152}},\
  \bibinfo {pages} {201102} (\bibinfo {year} {2020})}\BibitemShut {NoStop}%
\bibitem [{\citenamefont {Avron}(1998)}]{Avron}%
  \BibitemOpen
  \bibfield  {author} {\bibinfo {author} {\bibfnamefont {J.~E.}\ \bibnamefont
  {Avron}},\ }\bibfield  {title} {\enquote {\bibinfo {title} {Odd viscosity},}\
  }\href {\doibase 10.1023/A:1023084404080} {\bibfield  {journal} {\bibinfo
  {journal} {J. Stat. Phys.}\ }\textbf {\bibinfo {volume} {92}},\ \bibinfo
  {pages} {543--557} (\bibinfo {year} {1998})}\BibitemShut {NoStop}%
\bibitem [{\citenamefont {Han}\ \emph {et~al.}(2021)\citenamefont {Han},
  \citenamefont {Fruchart}, \citenamefont {Scheibner}, \citenamefont
  {Vaikuntanathan}, \citenamefont {{d}e Pablo},\ and\ \citenamefont
  {Vitelli}}]{Vitellis_pretty_pictures}%
  \BibitemOpen
  \bibfield  {author} {\bibinfo {author} {\bibfnamefont {M.}~\bibnamefont
  {Han}}, \bibinfo {author} {\bibfnamefont {M.}~\bibnamefont {Fruchart}},
  \bibinfo {author} {\bibfnamefont {C.}~\bibnamefont {Scheibner}}, \bibinfo
  {author} {\bibfnamefont {S.}~\bibnamefont {Vaikuntanathan}}, \bibinfo
  {author} {\bibfnamefont {J.~J.}\ \bibnamefont {{d}e Pablo}}, \ and\ \bibinfo
  {author} {\bibfnamefont {V.}~\bibnamefont {Vitelli}},\ }\bibfield  {title}
  {\enquote {\bibinfo {title} {Fluctuating hydrodynamics of chiral active
  fluids},}\ }\href {\doibase 10.1038/s41567-021-01360-7} {\bibfield  {journal}
  {\bibinfo  {journal} {Nat. Phys.}\ }\textbf {\bibinfo {volume} {17}},\
  \bibinfo {pages} {1260--1269} (\bibinfo {year} {2021})}\BibitemShut {NoStop}%
\bibitem [{\citenamefont {Weisenborn}\ and\ \citenamefont
  {Mazur}(1984)}]{2d_Stokes}%
  \BibitemOpen
  \bibfield  {author} {\bibinfo {author} {\bibfnamefont {A.~J.}\ \bibnamefont
  {Weisenborn}}\ and\ \bibinfo {author} {\bibfnamefont {P.}~\bibnamefont
  {Mazur}},\ }\bibfield  {title} {\enquote {\bibinfo {title} {The {O}seen drag
  on a circular cylinder revisited},}\ }\href {\doibase
  10.1016/0378-4371(84)90111-0} {\bibfield  {journal} {\bibinfo  {journal}
  {Physica A}\ }\textbf {\bibinfo {volume} {123}},\ \bibinfo {pages} {191--208}
  (\bibinfo {year} {1984})}\BibitemShut {NoStop}%
\bibitem [{\citenamefont {Hargus}\ and\ \citenamefont
  {Mandadapu}(2022)}]{Cory_github}%
  \BibitemOpen
  \bibfield  {author} {\bibinfo {author} {\bibfnamefont {C.}~\bibnamefont
  {Hargus}}\ and\ \bibinfo {author} {\bibfnamefont {K.~K.}\ \bibnamefont
  {Mandadapu}},\ }\href@noop {} {\enquote {\bibinfo {title} {Molecular
  simulations of active matter},}\ }\bibinfo {howpublished}
  {https://github.com/mandadapu-group/active-matter} (\bibinfo {year}
  {2022})\BibitemShut {NoStop}%
\end{thebibliography}%

\end{document}


\title{Supplemental Material for ``Odd mobility of a passive tracer in a chiral active fluid''}

\author{Anthony R. Poggioli}
\email{arpoggioli@berkeley.edu}
\affiliation{Department of Chemistry, University of California, Berkeley}
\affiliation{Kavli Energy NanoScience Institute, Berkeley, California}

\author{David T. Limmer}
\email{dlimmer@berkeley.edu}
\affiliation{Department of Chemistry, University of California, Berkeley}
\affiliation{Kavli Energy NanoScience Institute, Berkeley, California}
\affiliation{Materials Science Division, Lawrence Berkeley National Laboratory}
\affiliation{Chemical Science Division, Lawrence Berkeley National Laboratory}

\date{\today}

\maketitle

\section{Simulation Details}
\label{sec:simulation}

All simulations consisted of $N_{\rm tracer} = 4$ passive tracer particles and $N_{\rm mono} = 4996$ monomers comprising $N_{\rm dimer} = N_{\rm mono}/2 = 2498$ harmonically bonded dimers. The side-length of the two-dimensional periodic simulation box was adjusted to obtain the desired densities. For monomer densities of $\bar{n} = 0.2 , \, 0.4,$ and $0.8\, \sigma^{-2}$, the side-lengths were set to $L \approx 158.11, \, 111.80$, and $79.057 \, \sigma$, respectively, where $\sigma$ is the monomer diameter. The monomers are evolved in time according to the underdamped Langevin equation of motion
\begin{equation}
m \dot{\boldsymbol v}_i^{\alpha} = - \gamma {\boldsymbol v}_i^{\alpha} + {\boldsymbol F}_{c,i}^{\alpha} + {\boldsymbol F}_{a,i}^{\alpha} + \sqrt{2 \gamma k_B T} {\boldsymbol N}_i^{\alpha}(t),
\end{equation}
where $i \in \{1,N_{\rm dimer}\}$ indexes the dimer, $\alpha \in \{1,2\}$ indexes the individual monomer on dimer $i$, $m$ is the monomer mass, ${\boldsymbol v}_i^{\alpha} \equiv \dot{\boldsymbol x}_i^{\alpha}$ is the monomer velocity, ${\boldsymbol x}_i^{\alpha}$ is the monomer position, $\gamma$ is a friction coefficient describing dissipation into the underlying bath, ${\boldsymbol F}_{c,i}^{\alpha}$ is the sum of all conservative forces acting on the monomer, ${\boldsymbol F}_{a,i}^{\alpha}$ is the active force acting on the monomer, and ${\boldsymbol N}_i^{\alpha}(t)$ is a Gaussian white noise term satisfying
\begin{equation}
\left\langle {\boldsymbol N}_i^{\alpha}(t) \right\rangle = 0
\label{eqn:noise_mean}
\end{equation}
and
\begin{equation}
\left\langle {\boldsymbol N}_i^{\alpha}(t) \otimes {\boldsymbol N}_j^{\beta}(t') \right\rangle = \delta_{ij} \delta_{\alpha \beta} \delta (t - t') {\boldsymbol 1}.
\label{eqn:noise_correlation}
\end{equation}
This term represents noise from interactions with the underlying bath, and its prefactor $\sqrt{2 \gamma k_B T}$ is related to the friction coefficient $\gamma$ by the assumption of local detailed balance and the consequent imposition of a fluctuation-dissipation relation.

Physically, the assumption of Langevin dynamics corresponds to scenario that monomer and tracer particles are suspended on a viscous, quiescent, passive bath at a fixed temperature that injects and dissipates energy such that a local detailed balance is maintained between each degree of freedom and the bath. We are neglecting any hydrodynamics of the underlying bath, including global flows induced by the motion of the suspended dimer fluid, as well as any memory effects or hydrodynamic interactions induced by flows generated around the individual monomers or tracer particles. The model does not conserve momentum as it is lost by dissipation to the fluid bath and would therefore be classified as a ``dry model'' according to the classification scheme of Ref. \cite{joanny}.

The total conservative force ${\boldsymbol F}_{c,i}^{\alpha}$ acting on a given monomer is the sum of two terms: the harmonic bond between a given monomer and its intra-dimer pair and the Weeks-Chandler-Andersen (WCA) \citep{WCA} repulsive interaction between the given monomer and any other {\it non-bonded} monomers within the WCA cutoff, $r_{\rm WCA} = 2^{1/6} \sigma$. The harmonic and WCA potentials are given by, respectively,
\begin{equation}
u_{\rm harmonic} \left( d_i; k, d_0 \right) = k \left( d_i - d_0 \right)^2,
\label{eqn:harmonic}
\end{equation}
and
\begin{equation}
\begin{aligned}
u_{\rm WCA} &\left( r_{ij}^{\alpha \beta}; \sigma, \epsilon \right) = \\
&\!\begin{cases}
4 \epsilon \left[ \left( \frac{\sigma}{r_{ij}^{\alpha \beta}} \right)^{12} - \left( \frac{\sigma}{r_{ij}^{\alpha \beta}} \right)^6 \right] + \epsilon, & r_{ij}^{\alpha \beta} < r_{\rm WCA} \\
0, & r_{ij}^{\alpha \beta} \geq r_{\rm WCA}.
\end{cases}
\end{aligned}
\label{eqn:WCA}
\end{equation}
In the above, $k$ is a spring stiffness constant, set equal to $100 \, \epsilon \, \sigma^{-2}$ in all of our simulations, $d_i \equiv \left| {\boldsymbol d}_i \right| \equiv \left| {\boldsymbol x}_i^1 - {\boldsymbol x}_i^2 \right|$ is the instantaneous bond length of dimer $i$, ${\boldsymbol d}_i$ is the instantaneous bond vector, $d_0 = \sigma$ is the equilibrium bond length, set equal to the monomer diameter $\sigma$ in all of our simulations, $\epsilon$ is the WCA energy parameter, and $r_{ij}^{\alpha \beta} \equiv \left| {\boldsymbol x}_i^{\alpha} - {\boldsymbol x}_j^{\beta} \right|$ is the magnitude of the distance between two non-bonded monomers $i$-$\alpha$ and $j$-$\beta$. We therefore have ${\boldsymbol F}_{c,i}^{\alpha} = - \nabla_{{\boldsymbol x}_i^{\alpha}} \left( u_{\rm harmonic} + u_{\rm WCA} \right)$.

Each monomer experiences an active force of fixed magnitude $F_a$ directed orthogonal to the instantaneous bond vector ${\boldsymbol d}_i$ associated with the dimer of which it is a member. This results in an instantaneous active torque of magnitude $F_a d_i(t)$ acting on dimer $i$ and an average active torque of magnitude $F_a d_0$, given that the density is not so large that it results in an average bond length $\left\langle d_i \right\rangle < d_0$. The direction of the active force is such that $F_a > 0$ corresponds to a counterclockwise active torque. That is, the active forces on monomers $1$ and $2$ of dimer $i$ are related to the bond vector by
\begin{equation}
{\boldsymbol F}_{a,i}^1 = - {\boldsymbol F}_{a,i}^2 = - F_a {\boldsymbol \epsilon} \cdot \hat{\boldsymbol d}_i
\label{eqn:Fai12}
\end{equation}
where $\hat{\boldsymbol d}_i \equiv {\boldsymbol d}_i / \left| {\boldsymbol d}_i \right|$ is the unit vector pointing along the instantaneous bond of dimer $i$, and ${\boldsymbol \epsilon} \equiv \hat{\boldsymbol x} \otimes \hat{\boldsymbol y} - \hat{\boldsymbol y} \otimes \hat{\boldsymbol x}$ is the Levi-Civita tensor. Action of $\boldsymbol \epsilon$ on a vector results in clockwise rotation of that vector by an angle $\pi/2$.

This active dimer fluid model is identical to that introduced in \cite{Cory_MD}, and we employ this model here because it is two-dimensional and breaks parity and time-reversal symmetry and is therefore a minimal model of a chiral active fluid that exhibits odd viscosity \cite{Avron}. The underlying odd viscosity of the fluid is crucial to observe odd mobility of a passive tracer. Additionally, it is a simpler model than the frictional disk models employed in, for example, Ref. \cite{Vitellis_pretty_pictures}, because we only have to consider bonded point particle, rather than rigid body, dynamics.

The tracer particles are propagated according to the Langevin equation of motion
\begin{equation}
m \dot{\boldsymbol V}_I = - \gamma {\boldsymbol V}_I + {\boldsymbol F}_{{\rm WCA},I} + \sqrt{2 \gamma k_B T} {\boldsymbol N}_I (t),
\label{eqn:tracer_eom}
\end{equation}
where $I \in \left\{ 1, N_{\rm tracer} \right\}$ indexes the tracer particles, ${\boldsymbol V}_I = \dot{\boldsymbol X}_I$ is the velocity of tracer $I$, ${\boldsymbol X}_I$ is the position of tracer $I$, and ${\boldsymbol N}_I(t)$ is again a delta-correlated Gaussian white noise characterizing interactions with the underlying bath. ${\boldsymbol F}_{{\rm WCA},I}$ is the sum of all WCA interactions between surrounding monomers and the passive tracer and is given by
\begin{equation}
{\boldsymbol F}_{{\rm WCA},I} = - \nabla_{{\boldsymbol X}_I} \sum_{i\alpha} u_{\rm WCA} \left( \left| {\boldsymbol x}_i^{\alpha} - {\boldsymbol X}_I \right|; \sigma_{\rm tracer}, \epsilon \right).
\label{eqn:FWCA}
\end{equation}

For simplicity, the mass $m$ and friction coefficient $\gamma$ are taken to be the same for the tracer particles as they are for the monomers. In all of our simulations except those where we explicitly examine the dependence of the ratio of odd to even mobilities $\mu_o/\mu_e$ on tracer diameter $\sigma_{\rm tracer}$ (Fig. 2C, main paper), we take $\sigma_{\rm tracer} = \sigma$. In this case, since the friction coefficient is anticipated to be a function of the particle radius for spherically symmetric particles, we anticipate $\gamma$ to be the same for both the tracer and the monomers. For values of $\sigma_{\rm tracer} > \sigma$, the tracer friction coefficient would be larger than the monomer friction coefficient. However, in two-dimensions, the dependence of friction coefficient on particle radius is logarithmic and therefore relatively weak \cite{2d_Stokes}. Any dependence of $\gamma$ on the tracer diameter is therefore expected to be weak over the range of values of $\sigma_{\rm tracer}$ examined here and to affect only the quantitative values of mobilities and not the qualitative behavior of the ratio $\mu_o/\mu_e$ that we are interested in.

All simulations are conducted in LAMMPS. The active forces on the monomers are implemented via a custom code available at \cite{Cory_github}. We use Lennard-Jones units in all of our simulations, meaning that all quantities are measured in combinations of $m$, $\sigma$, and $\epsilon$ -- respectively, the monomer mass and diameter and the energy parameter characterizing the monomer-monomer WCA interaction. This is equivalent to setting $m = \sigma = \epsilon = 1$ in our simulations. We likewise take $\gamma = 1 \, \sqrt{m \epsilon \, \sigma^{-2}}$ in all of our simulations. The energy parameter characterizing the monomer-tracer interaction is also set to one, and the length parameter is set to $\sigma_{\rm tracer}$, defining precisely what is meant by the tracer diameter. All simulations are conducted at a temperature $k_B T / \epsilon = 1$. The time steps used in simulation were on the order of $10^{-3} \, \sqrt{m \sigma^2 \epsilon^{-1}}$, and an initial configuration is generated for each value of the density, activity, and tracer size from an initial grid configuration of particles over $\sim \mathcal{O} \left( 10^7 \right)$ time steps. Ensemble averages were evaluated over $\mathcal{O} \left( 10 - 100 \right)$ trajectories of lengths $\sim \mathcal{O} \left( 10^6 \right)$ time steps with randomized seeds for the Gaussian noise until sufficient convergence was obtained.

\section{Total Correlation Functions Determining Even and Odd Mobility Responses}
\label{sec:total_correlation}

\begin{figure}[t]
	\centerline{\includegraphics[width=8.5cm]{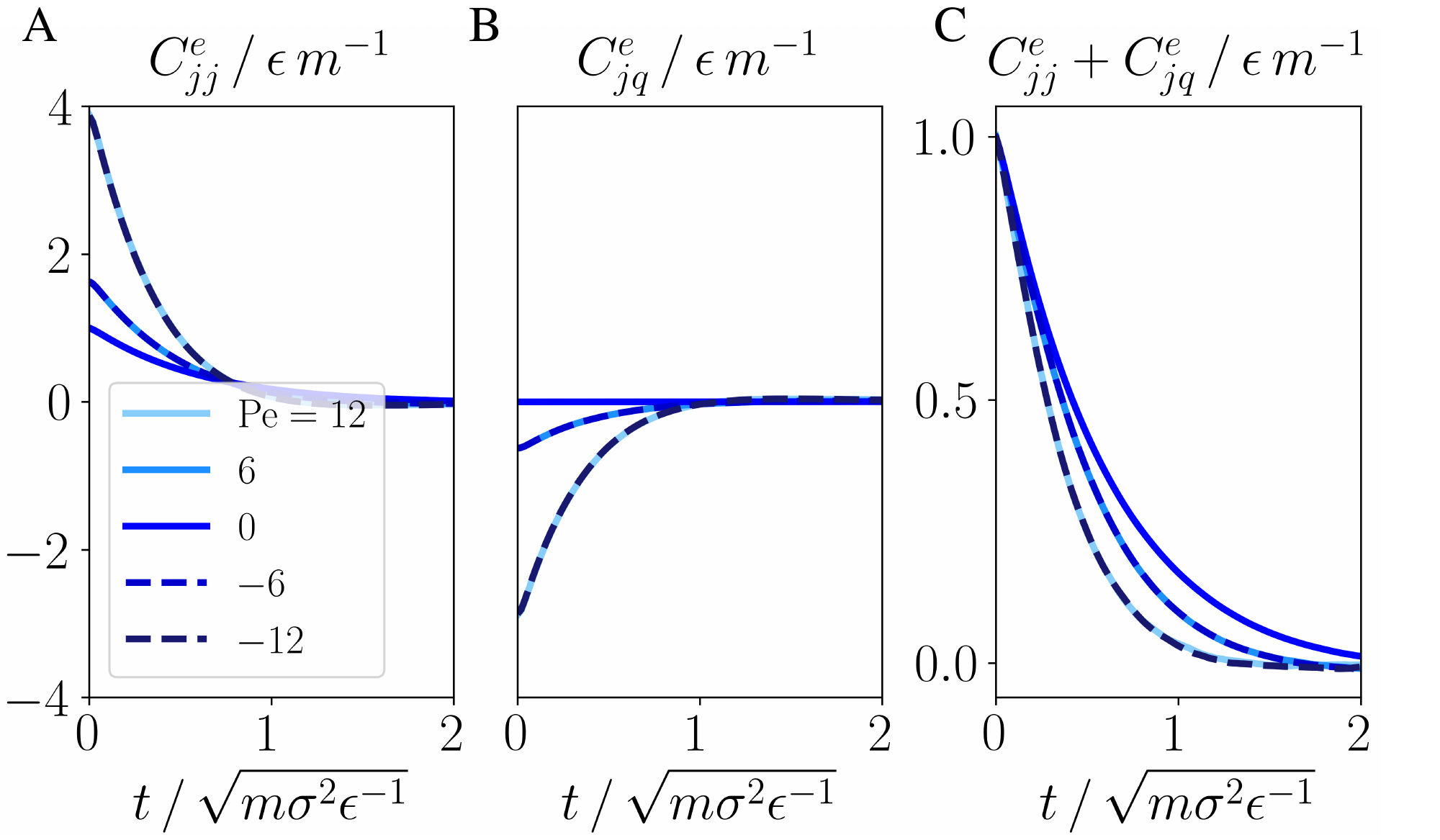}}
    \caption{Thermodynamic, frenetic, and total correlation functions determining the even mobility for different activities. A) Thermodynamic (current-current) contribution to the even mobility correlation function for different activities. B) Frenetic (current-frenesy) contribution to the even mobility correlation function. C) Total correlation functions determining the even mobilities, obtained as the sum of the thermodynamic and frenetic contributions.}
\label{fig:corr}
\end{figure}

In Figs. 1A and B of the main text, we found that increasing activity resulted in a substantial change in the variances associated with the thermodynamic and frenetic contributions to the even mobility correlation function; this is also indicated in Figs. \ref{fig:corr}A and B here. We show in Fig. \ref{fig:corr}C the total even mobility correlation function, obtained as the sum of the thermodynamic and frenetic contributions. Surprisingly, we find that the changes in the thermodynamic and frenetic variances exactly cancel such that the overall variance is unity for all of the activities examined. This is the same value that is guaranteed by equipartition for the variance in equilibrium. The origin of this cancellation and apparent ``effective equipartition'' is beyond the scope of this article and the subject of future research.

\section{Role of Temporal Boundary Term in Mobility Response}
\label{sec:unsteady}

In the following discussion, as in the main text, we note that the dilute tracer concentration ($\bar{n}_{\rm tracer} \equiv N_{\rm tracer}/L^2 \sim \mathcal{O} ( 10^{-4} \, \sigma^{-2} )$) allows us to neglect tracer-tracer interactions and therefore drop the tracer subscript $I$. In the main text, we define the time-extensive frenesy as ${\boldsymbol Q} \equiv - \int_0^{t_N} {\rm d} t \, {\boldsymbol F}_{\rm WCA} / \gamma$; however, from the Onsager-Machlup form of the path action (main text Eq. 5), we find ${\boldsymbol Q} = \int_0^{t_N} {\rm d} t \, \left( m \dot{\boldsymbol V} - {\boldsymbol F}_{\rm WCA} \right) / \gamma$. The stated form of the frenesy is obtained by neglecting the temporal boundary term ${\boldsymbol Q}_{\rm boundary} \equiv \int_0^{t_N} {\rm d} t \, m \dot{\boldsymbol V} / \gamma = (m/\gamma) \left[ {\boldsymbol V}(t_N) - {\boldsymbol V}(0) \right]$.

We can show that this boundary term must vanish in the limit $t_N \rightarrow \infty$, the limit in which our linear response results become valid. Including the additional additive boundary contribution to the frenesy ${\boldsymbol Q}_{\rm boundary}$ in the linear response relationship for the mobility tensor (main text Eq. 9), we find an additional additive contribution to the mobility tensor from the boundary term given by
\begin{equation}
\begin{split}
{\boldsymbol \mu}_{\rm boundary} &= \frac{\beta}{2 t_N} \left\langle \delta {\boldsymbol J} \otimes \delta {\boldsymbol Q}_{\rm boundary} \right\rangle_0 \\
&= \frac{\beta m}{2 \gamma t_N} \int_0^{t_N} {\rm d} t \, \left\langle {\boldsymbol V}(t) \otimes \left[ {\boldsymbol V}(t_N) - {\boldsymbol V}(0) \right] \right\rangle_0,
\end{split}
\label{eqn:mu_boundary}
\end{equation}
where in the second equality we have inserted the definitions of ${\boldsymbol Q}_{\rm boundary}$ and ${\boldsymbol J} \equiv \int_0^{t_N} {\rm d} t \, {\boldsymbol V}$ and noted that $\left\langle {\boldsymbol V} \right\rangle_0 = 0$. We can rewrite the first term in the second equality as
\[
\begin{split}
\int_0^{t_N} {\rm d} t \, \left\langle {\boldsymbol V}(t) \otimes {\boldsymbol V}(t_N) \right\rangle_0 &= \int_0^{t_N} {\rm d} t \, \left\langle {\boldsymbol V}(0) \otimes {\boldsymbol V}(t_N - t) \right\rangle_0 \\
&= \int_0^{t_N} {\rm d} t \, \left\langle {\boldsymbol V}(0) \otimes {\boldsymbol V}(t) \right\rangle_0.
\end{split}
\]
The first equality follows from time-translation invariance, and the second from a change of integration variable. The boundary contribution to the mobility tensor then becomes
\begin{equation}
\begin{split}
{\boldsymbol \mu}_{\rm boundary} &= \frac{\beta m}{2 \gamma t_N} \int_0^{t_N} {\rm d} t \, \langle {\boldsymbol V}(0) \otimes {\boldsymbol V}(t) \\
&\quad\quad\quad\quad\quad\quad\quad\quad\quad - {\boldsymbol V}(t) \otimes {\boldsymbol V}(0) \rangle_0 \\
&= \frac{\beta m}{2 \gamma t_N} \left\langle {\boldsymbol V}(0) \otimes {\boldsymbol J} - {\boldsymbol J} \otimes {\boldsymbol V}(0) \right\rangle_0.
\label{eqn:mu_boundary2}
\end{split}
\end{equation}

The fact that the thermodynamic contribution to the mobility tensor ${\boldsymbol \mu}_{\rm thermo} = (\beta/2t_N) \left\langle \delta {\boldsymbol J} \otimes \delta {\boldsymbol J} \right\rangle_0$ is finite implies that $\left\langle \delta {\boldsymbol J} \otimes \delta {\boldsymbol J} \right\rangle_0 \sim t_N$ in the long time limit, and therefore $\left\langle {\boldsymbol V}(0) \otimes {\boldsymbol J} \right\rangle_0/t_N$ and $\left\langle {\boldsymbol J} \otimes {\boldsymbol V}(0) \right\rangle_0/t_N$ must tend to zero as $t_N \rightarrow \infty$. From this fact and Eq. \ref{eqn:mu_boundary2} we conclude that ${\boldsymbol \mu}_{\rm boundary} \rightarrow 0$; that is, calculations of the mobility coefficients must converge to the same value whether they include the unsteady term $\int m \dot{\boldsymbol V}/\gamma$ in the frenetic contribution or not.

\begin{figure}[h!]
	\centerline{\includegraphics[width=8.5cm]{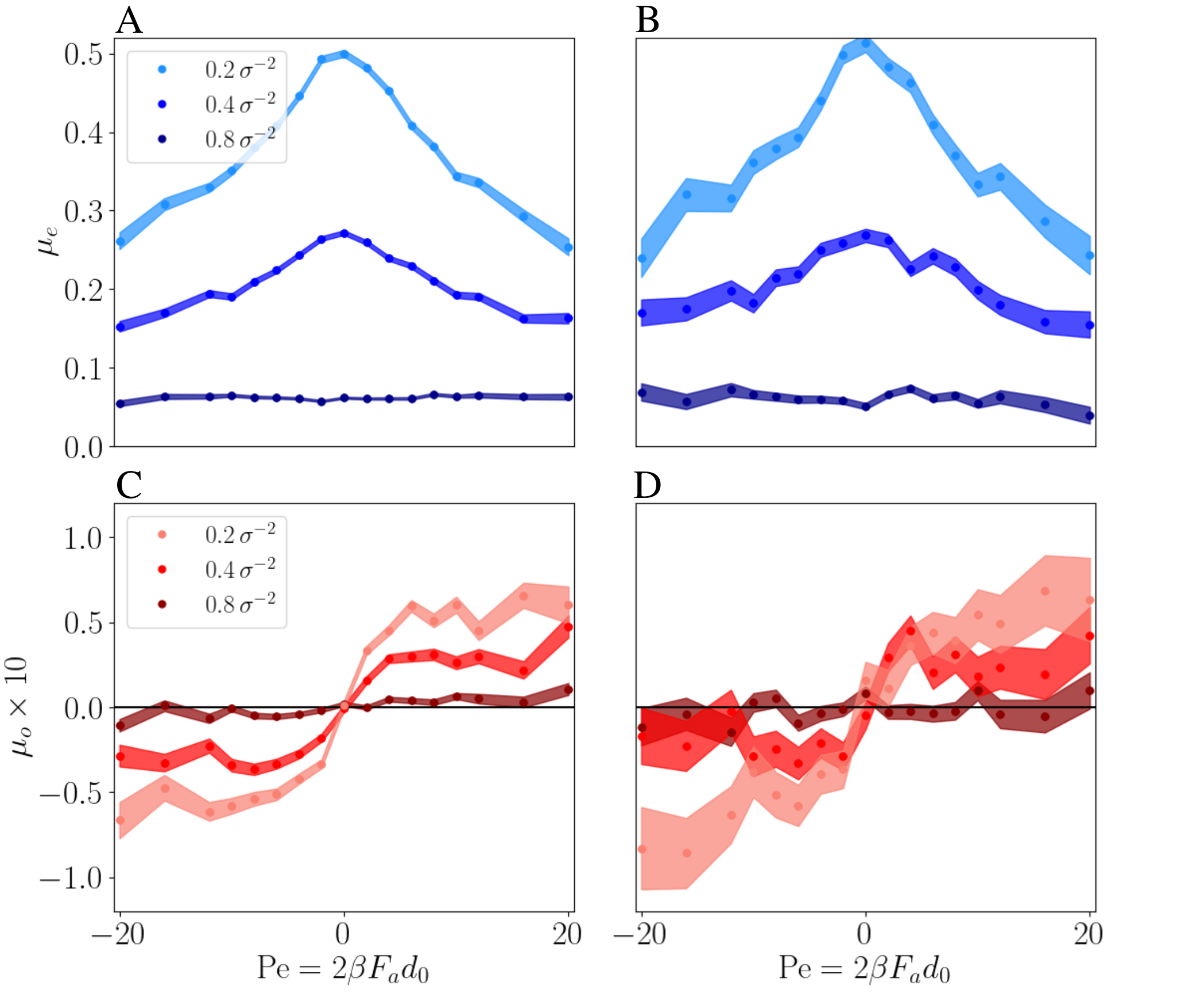}}
    \caption{Estimates of even and odd mobilities from nonequilibrium linear response theory without and with the temporal boundary contribution to the frenesy ${\boldsymbol Q}_{\rm boundary}$ for several different densities. A and B) Even mobility curves calculated without (A) and with (B) inclusion of the boundary frenesy. C and D) Odd mobility curves without (C) and with (D) the boundary frenesy contribution.}
\label{fig:unsteady}
\end{figure}

In numerical simulations it is necessary to choose a value of $t_N$ that is finite, though much larger than the integral relaxation times associated with the underlying correlation functions. We should therefore also confirm that our results are sufficiently converged such that the unsteady term is also irrelevant for the finite value of $t_N = 30 \, \sqrt{m \sigma^2 \, \epsilon^{-1}}$ we have used to calculate our response estimates of the mobility. This is done in Fig. \ref{fig:unsteady}, where we compare our estimates of the even and odd mobilities without and with the unsteady frenetic contribution. We see that there is no statistically significant difference in the mobility estimates, and that the only effect of including the unsteady term is to greatly increase the noise in our estimates. Therefore, we conclude that we are justified in neglecting the unsteady frenetic contribution to the mobility response.

\bibliography{mobility_ref}